\documentclass[structabstract]{aa}  
 \usepackage{longtable}
\usepackage{epsfig}
\usepackage{graphicx}
\usepackage{txfonts}
\usepackage{amssymb}
\usepackage{natbib}
\bibpunct{(}{)}{;}{a}{}{,}

\newcommand{\rxte}{{\it RXTE}} 
\newcommand{\integral}{{\it INTEGRAL}}
\newcommand{\rmd}{{\rm d}}
\newcommand{\msun}{{\rm M}_{\sun}}
\newcommand{\rg}{r_\mathrm{s}}
\newcommand{\Deltag}{\Delta_\mathrm{g}}
\newcommand{\beq}{\begin{eqnarray}}
\newcommand{\eeq}{\end{eqnarray}}
\newcommand{\be}{\begin{equation}}
\newcommand{\ee}{\end{equation}}
\newcommand{\bez}{\begin{eqnarray*}}
\newcommand{\eez}{\end{eqnarray*}}
\newcommand{\vecn}{\mbox{\boldmath $n$}}
\newcommand{\vecr}{\mbox{\boldmath $r$}}
\newcommand{\ani}{h}

\newcommand{\imin}{i_{\min}}
 
\newcommand{\thetas}{\theta_\mathrm{s}}
\newcommand{\thetamax}{\theta_\mathrm{m}}
\newcommand{\phis}{\varphi_\mathrm{s}}
\newcommand{\alphas}{\alpha_\mathrm{s}}

\begin{document}
\titlerunning{Light curves of X-ray pulsars}
   \title{Constraining compactness and magnetic field geometry of X-ray pulsars from 
   the statistics of their pulse profiles}

   \author{M. Annala
          \and
          J. Poutanen
}

   \institute{Astronomy Division, Department of Physics,  P.O. Box 3000,  
FIN-90014 University of Oulu, Finland\\
              \email{marja.annala@oulu.fi, juri.poutanen@oulu.fi}
             }

   \date{ } 

 
  \abstract
{The light curves observed from X-ray pulsars and magnetars reflect the radiation emission pattern, the geometry 
of the magnetic field, and the neutron star compactness.   }  
{We study the statistics of X-ray pulse profiles in order to constrain the neutron star compactness and the magnetic field geometry.}
{We collect the data for 124 X-ray pulsars, which are mainly in high-mass X-ray binary systems, and classify their 
pulse profiles according to the number of observed peaks seen during one spin period, dividing them into two classes,
single- and double-peaked.  We find that the pulsars are distributed about equally between both groups.
We also compute  the probabilities predicted by the theoretical models of two antipodal point-like spots that emit radiation according 
to the pencil-like emission patterns. These are then compared to the  observed fraction of pulsars in the two classes. 
}
{Assuming a blackbody emission pattern, it is possible to constrain the neutron star compactness 
if the magnetic dipole has arbitrary inclinations to the pulsar rotational axis. 
More realistic pencil-beam patterns predict that  79\%   of the pulsars  are double-peaked independently 
of their compactness. The theoretical predictions can be made consistent with the data if the magnetic dipole inclination 
to the rotational axis has an upper limit of 40\degr $\pm$4\degr.
We also discuss the effect of   limited sensitivity of the X-ray instruments to detect weak pulses, 
which lowers the number of detected double-peaked profiles and makes 
the theoretical predictions to be consistent with the data even if the  magnetic dipole does have random inclinations. 
This shows that the statistics of pulse profiles does not allow us to constrain the neutron star compactness. 
In contrast to the previous claims by Bulik et al. (2003), the data also do not require the magnetic
inclination to be confined in a narrow interval. }
{} 
 

\maketitle
%
%

\section{Introduction}

X-ray pulsars have been discovered in the 1970s \citep{1971ApJ...167L..67G} and have served as laboratories to study the neutron star (NS) physics since then. Most of them are members of binary systems and accrete matter through wind or via disk from a high-mass companion. Because of a large magnetic field strength (typically $10^{12}$ G) the material is channeled onto small spots at the magnetic poles. Here the relativistically moving plasma is decelerated in a radiative shock near the surface and this sub-sonically settling plasma radiates in the X-ray band \citep[see e.g.][]{{1973ApJ...179..585D},{1976MNRAS.175..395B}}. Pulsations are observed if the magnetic field is inclined relative to the rotation axis. 
Even stronger field pulsars (magnetars) have been discovered recently, which operate by dissipating magnetic energy \citep[see e.g. review by][]{2008A&ARv..15..225M}. Studies of the pulse profiles of individual  pulsars  allow one to constrain the emission pattern of the hotspots (or accretion columns) at the NS surface as well as the geometry of the magnetic field \citep[see e.g.][]{{1980A&A....90...26Y},{1995ApJ...444..405B},{1996ApJ...467..794K}}. 

The number of known pulsars in the Milky Way and the nearby Small and Large Magellanic clouds is already above a hundred. The quality of the data is also improving because of the sensitive X-ray/gamma-ray observatories such as \rxte\ and \integral\ and the long observing times.  The first pulsars have already been discovered in M31 \citep{2005ApJ...634..314T} and in even more distant galaxies \citep{2007ApJ...663..487T,2008MNRAS.387L..36T}. A large number of pulsars allows us to use a statistical approach to constrain the NS parameters. 
Because of the gravitational light bending, the more compact the star, the larger the fraction of the NS surface that is visible to an observer at all times. 
This increases the probability to observe two radiative poles in one rotational period and affects the relative fraction of the single- and double-peaked pulse profiles.
The second important parameter that affects the pulse profile is the positions of the hotspots relative to the rotational axis (which are defined in the simplest case by the inclination of the magnetic dipole). 

For any reasonable NS parameters one expects that both poles are visible in a large majority of pulsars. 
However, the number of observed single- and double-peaked profiles is not so different.  
This was already noticed by \citet{1981AAA...102...97W} and later by \citet[][ B03 hereafter]{2003AAA...404.1023B}.  
This discrepancy can be explained if the inclination between the magnetic dipole and rotational axis is not randomly distributed, but if there is a strong bias towards alignment. 
In the present study,  we consider a sample of 124 pulsars with better quality data than were available before. 
We study in detail the statistics of double-peak profiles for various pencil-like emission pattern. We also 
discuss the effect of the detection threshold that can significantly affect the observed fraction. 
We then derive useful analytical formulae that describe the probabilities of observing certain types of pulsars. 
And finally we  compare our theoretical model to the data.

\begin{table} 
\caption[]{Light curve classification of 60 X-ray pulsars observed at energies above 10 keV.}
\begin{tabular}{lcccc}
\hline\hline
Name & $P_{\rm spin}$\tablefootmark{a}  	&  Pulses\tablefootmark{b}  &	B03\tablefootmark{c} & Ref. \\ %
\hline
 &  & &  & \\
 RX J0051.8--7310 		& 16.6&1 && 1 \\
RX J0052.1--7319 	 	& 15.4 & 2 & 2 & 2\\
XTE-SMC95 			& 95	&1 & & 3 \\
SMC X-2 	 		& 2.37 & 1 & 1 & 4 \\
XTE J0055--724  		& 59	& 1 & 1 & 5 \\
SMC X-1 	 		& 0.71 &2&2& 6 \\
RX J0117.6--7330 		& 22	& 2 & 2 & 7\\
2S 0114+650   		& 9828 &1& 1 & 8\\
4U 0115+63  		& 3.6  &1 & 1 & 9, 16  \\
4U 0142+614\tablefootmark{d} & 8.7 & 2 & 2 & 10 \\
RX J0146.9+6121    	& 1408 & 1 & 1 & 11 \\
V 0332+53 	 		& 4.37  & 2 & 1 & 12 \\
4U 0352+309 = X Per   	& 835   & 1 & 1 & 13 \\
EXO 053109--6609.2   	& 13.7  & 2 &    & 14 \\
LMC X-4            		& 13.5  & 1 & 1 & 15 \\
1A 0535+26   		& 105   & 2 & 2 & 16 \\
MXB 0656--072 		& 160   & 1 &    & 17 \\
4U 0728--25  		& 103   & 2 & 2 & 18 \\ 
RX J0812.4--3114   	& 31.9  & 2 & 2 & 19 \\ 
GS 0834--430       		& 12.3  & 2 & 2 & 16, 20 \\
Vela X-1 		  	& 283 & 2& 2 & 16, 21 \\
GRO J1008--57  	 	& 93.5 & 1 & 1 & 16 \\
1A 1118--616   		& 405& 1 & 1&16, 22 \\
Cen X-3 	 		& 4.82 & 2 &	2 &   16, 23 \\
1E 1145.1--6141    	& 297 & 2 & 1 &  16, 21 \\
4U 1145--619 		& 292 & 1 & 1 &         16 \\
GX 301--2   			& 681 & 2 & 2 & 16, 24 \\
GX 304--1   			& 272 & 1(flat) & 1 & 25 \\
2S 1417--624 	 	& 17.6 & 2 & 2 & 16  \\
4U 1538--52 	 	& 530 & 2 & 2 & 16, 21 \\
XTE J1543--568 		& 27.1 & 2 &    & 26  \\
SWIFT J1626.6--5156 	& 15.4 & 1 & & 27 \\
IGR J16358--4726 		& 228 &1 & & 28 \\
IGR J16393--4643 		& 912 &2 & & 29 \\
OAO 1657--415 	  	& 37.7 & 1 & 1 & 16 \\ 
1RXS J170849.0--400910\tablefootmark{d}  & 11.0  &1 &1 & 10  \\
GPS 1722--363  		& 414 & 1 & 1 & 30  \\  
AX J1749.1--2733 		& 132 &2 & & 31  \\
GRO J1750--27 	 	& 4.45 & 1 &	1 &   16,  32 \\
SAX J1802.7--2017 	& 140 &2 & & 33 \\
SGR 1806--20\tablefootmark{d}   & 7.47 &2 & & 34  \\
Sct X-1 		 	& 111 & 2 &	2 & 35 \\  
GS 1843+00 	 	& 29.5 & 2 & 2 &         36 \\
GS 1843--024  	 	& 94.3 & 1  & 1 &     37 \\
IGR J18483--0311 		& 21.1 & 2 & & 38  \\
XTE J1855--026  	 	& 361 & 1	& 1 & 39 \\
XTE J1858+034 	 	& 221 & 1 & 1 &    40  \\
SGR 1900+14\tablefootmark{d}  & 5.16 &2 & & 34 \\
4U 1901+03 		& 2.76 &2 & &41 \\
XTE J1906+09  	 	& 89 & 2 & 1 &      42 \\
4U 1907+09   		& 440 & 2 & 2 & 43 \\
4U 1908+075 		& 605 &2 & & 44 \\
XTE J1946+274     		& 15.8 & 2 & 2 &      45 \\
KS 1947+300 		& 18.7 & 1 &  1 & 46, 47 \\ 
SW J2000.6+3210 		& 1056 & 1 & & 48  \\
EXO 2030+375 	 	& 42 & 2 & 2 &       16,  49 \\
GRO J2058+42 	 	& 198 & 1 &	1 &    16  \\
SAX J2103.5+4545		& 359 & 1 & 1 & 50 \\
Cep X-4  		 	& 66.2 & 2 & 2 & 51  \\
1E 2259+586\tablefootmark{d}  & 6.98 & 2 & 2 &  52 \\
 &  & &  & \\
\hline
\end{tabular}
\label{pulsartable1}
\tablefoottext{a}{Pulsar spin period (s).}
\tablefoottext{b}{Number of pulses in the profile.}
\tablefoottext{c}{Classification by B03.}
\tablefoottext{d}{Magnetars.}
\end{table}

\begin{table}
\tablebib{(1)~\citet{2002ApJ...567L.129L}; (2)~\citet{2001ApJ...560..378F};
(3)~\citet{2002AA...385..464L}; (4)~\citet{2001ApJ...548L..41C}; 
(5)~\citet{1998AAA...338L..59S}; (6)~\citet{1993ApJ...410..328L}; 
(7)~\citet{1999ApJ...518L..99M}; (8)~\citet{2000ApJ...536..450H};
(9)~\citet{1999ApJ...523L..85S}; (10)~\citet{2006ApJ...645..556K};
(11)~\citet{2000AAA...354..567M}; (12)~\citet{2006MNRAS.371...19T};
(13)~\citet{1989ApJ...346..469R}; (14)~\citet{1998ApJ...498..831B}; 
(15)~\citet{1996ApJ...467..811W}; (16)~\citet{1997ApJS..113..367B};
(17)~\citet{2006AA...451..267M}; (18)~\citet{1997ApJ...489L..83C}; 
(19)~\citet{1999MNRAS.306...95R}; (20)~\citet{1992PASJ...44..641A};   
(21)~\citet{1995PhDT.......215M}; (22)~\citet{1994AAA...289..784C};
(23)~\citet{2000ApJ...530..429B}; (24)~\citet{1997ApJ...479..933K};
(25)~\citet{1977ApJ...216L..15M}; 
(26)~\citet{2001ApJ...553L.165I}; (27)~\citet{2008AA...485..797R};
(28)~\citet{2005AA...444..821L}; 
(29)~\citet{2006AA...447.1027B}; (30)~\citet{1989PASJ...41..473T}; 
(31)~\citet{2008MNRAS.386L..10K}; (32)~\citet{1997ApJ...488..831S};
(33)~\citet{2003ApJ...596L..63A}; (34)~\citet{2002ApJ...577..929G};
(35)~\citet{1991ApJ...370L..77K};
(36)~\citet{1990ApJ...356L..47K}; (37)~\citet{1999ApJ...517..449F};
(38)~\citet{2007AA...467..249S}; (39)~\citet{1999ApJ...517..956C};
(40)~\citet{1998A&A...337..815P}; (41)~\citet{2005ApJ...635.1217G};
(42)~\citet{2002ApJ...565.1150W}; (43)~\citet{1998ApJ...496..386I};
(44)~\citet{2004ApJ...617.1284L}; (45)~\citet{2003ApJ...584..996W};
(46)~\citet{2005AstL...31...88T}; (47)~\citet{1995ApJ...446..826C}; 
(48)~\citet{2009ApJ...699..892M};  (49)~\citet{1999ApJ...512..313S};
(50)~\citet{1998AAA...337L..25H}; (51)~\citet{1991ApJ...366L..19K}; 
(52)~\citet{1992PASJ...44....9I}.
}
\end{table}

\section{Data selection and classification of light curves}
\label{sec:data}

The light curves of X-ray pulsars can be classified according to the number of pulses per period. 
The observed pulse profiles tend to simplify with increasing energy, and the multiple-peaked profiles change into double- or single-peaked
\citep[see e.g.][]{1989PASJ...41....1N,1997ApJS..113..367B}.
In order to reduce possible effects of the photoelectric  absorption and the cyclotron lines, the pulse classification 
is done at the highest possible energies (typically above 10 keV). 
We use  the published light curves of X-ray pulsars from several sources. Therefore the data are inhomogeneous. A significant number of light curves are produced in an non-appropriate energy range, which makes it difficult to conclude anything about the number of poles visible to the observer and details of the beam shape.

\begin{table} 
\caption[]{Light curve classification of 64 X-ray pulsars observed at energies below 10 keV.}
\begin{tabular}{lcccc}  
\hline\hline
Name 	&  $P_{\rm spin}$\tablefootmark{a}  & Pulses\tablefootmark{b}  &	B03\tablefootmark{c} & Ref. \\
\hline 
 &  & &  & \\
 XMMU J004723.7--731226 	& 263 &1 &   & 1 	 \\
AX J0049--729   	         		& 74.7&1 &1 & 2 	 \\
AX J0049--732   			& 9.1  &1 &1 & 3 	\\
AX J0049.5--7323   		& 756 &1 &   & 4 	\\
2E 0050.1--7247   		& 8.9  &1 &1 & 5 	\\
RX J0051.3--7216 			&91 &2 & & 6	\\ 
AX J0051--733   			&323 	&1 &1 & 7	 \\
AX J0051.6--7311 			& 172 &1 &   & 8 	\\
SMC X-3 				& 7.8  &1 &   & 9 	\\ 
XTE J0052--725 			& 82.5&1 &   & 9	\\
XTE J0052--723 			& 4.78&2 &  & 10	\\
CXOU J005323.8--722715  	&138 &1 &   & 9 	\\  
RX J0053.8--7226   		&46.6&1&  & 11 	\\
XTE J0054--720   			&168&1 &   & 12	 \\
CXOU J005455.6--724510  	&500 &1 & & 9,13	\\
RX J0054.9--7226 			&59 &1 & & 14	\\ 
XMMU J005517.9--723853 	&702 &1 & & 13	\\
CXOU J005527.9--721058  	&34.1 &2 & & 9 	 \\
XMMU J005605.2--722200 	&140&2 & & 14	\\ 
AX J0057.4--7325 			&101&1 & & 15	\\
CXOU J005736.2--721934  	&563 &1 & & 9,18 	\\
RX J0057.8--7207 			&152&1 & & 14 	 \\
AX J0058--720 			&281&2 &2& 14,16 \\
1XMMU J005921.0--722317 	&202&1 & & 1	\\
RX J0059.2--7138   		&2.76&1 &1 & 17	\\
CXOU J010043.1--721134\tablefootmark{d}  &8.02&2 & & 1 	 \\
CXOU J010102.7--720658  	&304&1 & & 18  	\\
RX J0101.3--7211			&452&1 & & 14 \\
XTE J0103--728 			&6.85&2& & 19      \\ 
SAX J0103.2--7209   		&345 &1&1 & 20  	 \\
RX J0103.6--7201 			&1323&2 & & 21 	\\ 
J0105--721   			&3.34&1 &  & 5	 \\ 
XTE J0111.2--7317  		&30.9& 2 & 1 & 22   \\
RX J0440.9+4431			&202 & 1 & 1 & 23 \\ 
RX J0502.9--6626   		&4.06 & 2 & 1 & 24 \\
RX J0529.8--6556   		&69 & 1 & 1 & 25 \\
XMMU J053011.2--655122 	&272 &2 & & 26 \\
EXO 053109--6609.2 		&13.7 &2 & & 26 \\
1A 0538--66    			&0.069 &2 &	2 &  27 \\
1SAX J0544.1--710   		&96 & 2 & 1 & 28 \\
SAX J0635.2+0533			&0.034 &1 &	1 & 29 \\
RX J0648.1--4419   		&13.2 & 1& 1 & 30 \\
RX J0720.4--3125   		&8.39 & 1 & 1 & 31 \\
RX J1037.5--5647			&860 & 2 & 1 & 23 \\ 
1E 1048.1--5937\tablefootmark{d} &6.44 &1 & 1 & 32 \\
IGR J11215--5952 			&187 &2 & &33 \\
IGR J11435--6109 			&162 &2 & & 34 \\
2RXP J130159.6--635806 	&710 &1 & & 35  \\
1SAX J1324.4--6200   		&171 & 1 & 1 & 36 \\
1SAX J1452.8--5949   		&437 &1 & 1& 37 \\
2S 1553--54(2)    			&9.3 & 2 & 2 & 38 \\
IGR J16320--4751 			&1300 &2 & & 39 \\
IGR J16465--4507 			&228 &1 & & 40 \\
CXOU J164710.2--455216\tablefootmark{d} &10.6 &2 & & 41  \\
AX J170006--4157   		&715 & 1 & 1 & 42 \\
AX J1740.1--2847 			&730 &2 & & 43 \\
AX J1749.2--2725   		&220 & 1 & 1 & 44 \\
XTE J1810--197 			&5.54 &1 & &45 \\
AX 1820.5--1434   		&152 &2&  & 46 \\
XTE J1829--098 			&7.8   &1 & & 47 \\
AX J1841.0--0536 			&4.74 &1 (flat) & & 48 \\
1E 1841--045\tablefootmark{d}  &11.8 &1 & & 49 \\
AX J1845.0--0300\tablefootmark{d} &7.0 & 1 & 1 & 50 \\
SAX J2239.3+6116 		&1247 &1  & & 51 \\
 &  & &  & \\
 \hline
\end{tabular}
\label{pulsartable2}
\tablefoottext{a}{Pulsar spin period (s).}
\tablefoottext{b}{Number of pulses in the profile.}
\tablefoottext{c}{Classification by B03.}
\tablefoottext{d}{Magnetars.}
\end{table}
\begin{table} 
\tablebib{(1)~\citet{2004ApJ...609..133M}; (2)~\citet{1999PASJ...51..547Y}; 
(3)~\citet{2000PASJ...52L..63U}; (4)~\citet{2000PASJ...52L..73Y};
(5)~\citet{1997ApJ...484L.141I}; (6)~\citet{2000ApJS..128..491Y}; 
(7)~\citet{1999PASJ...51L..15I};  (8)~\citet{2000PASJ...52L..37Y}; 
(9)~\citet{2004MNRAS.353.1286E}; (10)~\citet{2003MNRAS.339..435L};
(11)~\citet{2005ApJS..161...96L}; (12)~\citet{2001PASJ...53L...9Y}; 
 (13)~\citet{2004AA...420L..19H}; (14)~\citet{2003AA...403..901S}; 
(15)~\citet{2000PASJ...52L..53Y}; (16)~\citet{1999PASJ...51L..21T};
(17)~\citet{2000PASJ...52..299K}; (18)~\citet{2003ApJ...584L..79M};
(19)~\citet{2008AA...484..451H}; (20)~\citet{2000ApJ...531L.131I};
(21)~\citet{2005AA...438..211H}; (22)~\citet{2000ApJ...539..191Y}; 
(23)~\citet{1999MNRAS.306..100R}; (24)~\citet{1995PASP..107..450S}; 
(25)~\citet{1997AAA...318..490H}; (26)~\citet{2003AA...406..471H};  
(27)~\citet{1982Natur.297..568S};
(28)~\citet{1998AAA...337..772C}; (29)~\citet{2000ApJ...528L..25C}; 
(30)~\citet{1997ApJ...474L..53I}; (31)~\citet{2004AAA...419.1077H};
(32)~\citet{1986ApJ...305..814S}; (33)~\citet{2007AA...476.1307S}; 
(34)~\citet{2004ATel..362....1I}; (35)~\citet{2005MNRAS.364..455C};
(36)~\citet{1998AAA...339L..41A}; (37)~\citet{1999AAA...351L..33O}; 
(38)~\citet{1983ApJ...274..765K}; (39)~\citet{2005AA...433L..41L};
(40)~\citet{2005AA...444..821L}; (41)~\citet{2007ApJ...664..448I};  
(42)~\citet{1999ApJ...523L..65T}; (43)~\citet{2000PASJ...52.1141S};
(44)~\citet{1998ApJ...508..854T}; (45)~\citet{2005ApJ...618..874H}; 
(46)~\citet{1998ApJ...495..435K}; (47)~\citet{2007ApJ...669..579H}; 
 (48)~\citet{2001PASJ...53.1179B}; (49)~\citet{2003PASJ...55L..45M}; 
(50)~\citet{1998ApJ...503..843T}; (51)~\citet{2001AA...380L..26I}. }
\end{table}

\citet{2003AAA...404.1023B} have studied the profiles of 88 pulsars (not 89 because one of the sources, 1WGA J1958.2+3232, turned out to be an intermediate polar,~\citealt{2000A&A...354L..29N}), which have been divided into three different groups. The first group consisted of 46 pulsars, which were easy to classify. The second group consisted of 31 pulsars, which were difficult to classify and the third group had 11 pulsars for which there were no good quality light curves available in 2003. We have scanned the latest literature for the light curves from the same sources and also  added to our sample all newly discovered X-ray pulsars. All together the sample now consists of 124 X-ray pulsars. The  pulsars are placed in two different categories depending on the kind of the data: those which have the profiles observed above 10 keV  and those that do not.  The first category contains 60 pulsars and the second one 64 pulsars. Their data are given in Tables~\ref{pulsartable1} and \ref{pulsartable2}, respectively, together with their classification (i.e. the number of pulses observed). 
A similar classification from B03 (who found 38 double-peaked light curves out of 88 sources) is also shown for comparison. 
Only the pulsars residing in high-mass X-ray binaries and Be-transients as well as magnetars are included, while all disk-accreting systems residing in low-mass X-ray binaries are excluded. This is because the disk can seriously affect  the light propagation from the secondary pole to the observer \citep[e.g.][]{IP09}. 
 
A couple of the classified sources (e.g. 4U 1538--52) show profile changes in the energy ranges above 10 keV, and 
the secondary pulse was not observed in every energy band. 
The most probable reason for this is the cyclotron absorption.  
These sources were classified as double-peaked. In two sources, GX 304--1 and AX J1841.0--0536, pulsations become weak as the energy increased.
These sources were classified as single-peaked. 

In general, the probability of observing $M$ double-peaked pulsars out of $N$ sources, is given by the binomial distribution.  Because our data set is very large and the observed number of pulsars of both types is similar, we can use the normal distribution instead. 
The estimation of the  probability is $p=M/N$ and its error is $\sqrt{p(1-p)/N}$.
In our classification of all pulsars in Tables \ref{pulsartable1} and \ref{pulsartable2}  we found 55 double-peaked light curves out of 124 sources, which gives  the probability of observing double-peaked profiles 
\be \label{eq:p0_obs}
p_0  = 0.44 \pm 0.04. 
\ee
For those pulsars which have light curves above 10 keV (Table~\ref{pulsartable1}), we have 33 double-peaked pulsars out of 60 sources, which gives the corresponding probability of 
\be \label{eq:p1_obs}
p_1=0.55\pm 0.06. 
\ee
This is still consistent with $p_0$ within $2\sigma$. 
Excluding the magnetars from the list of pulsars would change the probabilities very little to 
$p'_0=0.43\pm0.05$ and $p'_1=0.52\pm 0.07$. 

\section{Modeling light curves of X-ray pulsars}

\subsection{Model setup}

In order to obtain some constraints on the (distribution of) NS parameters such as compactness, magnetic field 
inclination, and the emissivity pattern from the statistical data (such as the fraction of the double-peaked profiles), we 
need to make a set of simplifying assumptions regarding the NS and the emission.  
In most of the following discussion, we assume that the NSs have a dipole magnetic field and two antipodal, point-like radiating hotspots at the magnetic poles at the NS surface. We assume also that all NS have the same compactness and that the emission from pulsars is described by the same pencil-beam pattern. 
As we will see below, the assumption of the same emission pattern for all pulsars 
will not have much effect on the results and therefore could be relaxed.  
Thus the pulsars differ from each other by the observer inclination, magnetic field inclination, and 
possibly by the emission pattern. 
We now discuss our assumptions one by one.

\begin{enumerate}
\item Compactness. The gravitational light-bending effect depends only on the compactness, i.e. 
mass-to-radius ratio $M/R$, which we assume to be the same for all NS. 
This is reasonable, because the observed distribution of NS masses in radio pulsars is very narrow \citep{1999ApJ...512..288T,HPY07} and the accretion in high-mass systems could not provide a significant mass increase during the life time of the system. 
Below we will also show that the statistics of pulse profiles depends very little on the compactness for realistic emission patterns.

\item Point-like emission regions. 
The lower limit on the spot size  in accreting X-ray pulsars can be obtained by assuming that  the accreting matter is bound by the magnetic field lines intersecting the Alfv\'en radius \citep{1981AAA...102...97W}. For a pulsar with mass-accretion rate $\dot M =10^{-9} M_{\odot}$/year, magnetic field strength of $B= 10^{12}$ G, and typical NS mass $M=1.4 M_{\odot}$ and radius $R=10$ km, the typical size of the hotspots  is about 4 degrees. 
A study of the interchange instability and diffusion of plasma through the magnetic field gives a 
higher estimate of about 20 degrees \citep{1980ApJ...235.1016A}. 
Thus in any case, the radiating spot size is much smaller than the stellar radius 
and the pulse profiles will not be dramatically affected because of the strong gravitational bending. 
Increasing the spot size increases the probability to see the secondary spot, but unless the emission pattern 
is fan-like, the number of pulses will not be affected.  

For magnetars, the area of the thermally emitting region is only a few km$^2$ \citep{2002nsps.conf...29M}.
Although the nature of the non-thermal persistent emission of anomalous X-ray pulsars and soft-gamma ray repeaters above 10 keV   \citep{2004ApJ...613.1173K,2005A&A...433L..13M,2005A&A...433L...9M,2006ApJ...645..556K} is not known, it also can be produced in very localized regions close to magnetic poles \citep{2005ApJ...634..565T,2007ApJ...657..967B}. 
Thus the assumption of the spot-like regions seems reasonable and the corrections arising from a finite spot size are negligible.

\item Emission patterns. 
The exact geometry and the structure of the emission region in accreting X-ray pulsars is model-dependent and varies from  plane-parallel slabs \citep[see e.g.][]{{1981ApJ...251..288N},{1981ApJ...251..278N},{1986A&A...169..259K},{1985ApJ...298..147M}, {1985ApJ...299..138M}} to columns/mounds  \citep[see e.g.][and references therein]{{1976MNRAS.175..395B},{1991ApJ...367..575B}, {2001ApJ...563..289K},{2003ApJ...590..424K}, {2003ApJ...596.1131L},{2005ApJ...621L..45B}}. 
At high enough energies, where the pulse profiles are rather simple  \citep[see e.g.][]{1989PASJ...41....1N,1997ApJS..113..367B}, the effects of the cyclotron and photoelectric absorption are minimized. 
We parameterize the (possibly complicated) emission pattern with simple mathematical functions. 
For example, we take the pencil-like emission pattern with the surface flux given by $F\propto \cos^n\alpha$, where $\alpha$ is the inclination of the spot normal to the light of sight.  Detailed modeling of the pulse profiles of seven pulsars by \citet{1995MNRAS.277.1177L} showed that such a pattern with $n=2$--4 gives a good description of the data.  
These patterns also describe well theoretical dependences expected from a magnetized slab \citep{1985ApJ...299..138M} as was shown by 
\citet{1990MNRAS.242..188L}. 
The case $n=1$ corresponds to the blackbody-like emission. This emission pattern is probably not physical, but it is a useful starting point for discussing the pulsar classification scheme \citep{2002ApJ...566L..85B,PB06}.
We also consider an alternative beaming pattern $F\propto \cos\alpha (1 + h \cos\alpha) $, where $h>-1$ is a parameter.  
The case with $h<0$ would correspond to the scattering in an optically thin electron atmosphere associated with the accretion shock or heated NS surface layer \citep[see e.g.][]{VP04}, while $h>0$ resembles pencil-beam and 
is more appropriate for optically thick emission. 

Although the physics of the persistent emission from magnetars 
is  very different \citep{2005ApJ...634..565T,2007ApJ...657..967B}, the simplicity of their pulse profiles and their 
broad peaks argues in favor of broad emission beams, which can be represented by the assumed patterns.
 
We first assume that all pulsars can be described by the same emission pattern and 
then discuss the consequences of relaxing this assumption.  
We will show below that the statistics of pulse profiles 
depends very little on the actual emission pattern.

\item Antipodality. We assume that the hotspots are antipodal, even though many  pulsars 
show asymmetric profiles. The detailed modeling of the pulse profiles \citep{1995MNRAS.277.1177L,1996ApJ...467..794K} 
shows, however, that the displacement of the spots relative to the antipodal position vary from a few to about 10 degrees. 
This displacement, while causing the profile asymmetry, does not change the number of pulses, which is important 
for our analysis.  

\item Emission from the NS surface. 
Pulsars in Be-transient systems show time evolution in their pulse profiles during the outbursts related to the changing mass-accretion rate. This results either from variations in the emission pattern and/or changes in the accretion shock 
height. According to the recent cyclotron line measurements, the radiative region in accretion column does not extend higher than about 
7\% of the neutron star radius \citep{2006MNRAS.371...19T}. 
In magnetars the persistent emission most probably also originates from the stellar surface  \citep{2005ApJ...634..565T,2007ApJ...657..967B}. Therefore a simple model assuming the radiation is produced in the vicinity of the neutron star surface is justified.

\item Slow rotation. 
In principle, the stellar rotation could affect the profiles because of the effects of relativistic aberration and time delays \citep{PG03,PB06}, but typical pulsars in Be-transient systems and high-mass X-ray binaries as well as magnetars rotate too slowly for these effects to be important.  

\item Accretion disk. The accretion disk in strong magnetic field pulsars is  normally disrupted at a distance that is large compared to the NS radius, and therefore it does not affect the visibility of the radiative spots on the neutron star surface or the pulse profile.
These effects become important in NS in  low-mass X-ray binaries such as 
accreting millisecond X-ray pulsars \citep[see][]{P08AIP, IP09}, but we do not include these objects in our study.  
\end{enumerate}

\subsection{Pulsar classes and observed fluxes}

Let $\theta$ be  the angle between rotational and magnetic axes and $i$  the inclination of the rotational axis to the line-of-sight. Then the unit vector in  the observer's direction is $\vecn=(\sin i, 0, \cos i)$ and the unit vector in the direction of the primary spot from the NS center is $\vecr=(\sin\theta\cos\varphi,  \sin\theta\sin\varphi, \cos\theta)$ with $\varphi$ being the pulsar phase. As pulsar rotates the position of the spots relative to the observer changes. For the primary spot (closest to the observer)  
\begin{equation}\label{eq_rot}
\cos{\psi}= \vecn\cdot \vecr = \cos{\theta}\cos{i}+\sin{\theta}\sin{i}\cos{\varphi} .
\end{equation}
At $\varphi=0$ the primary spot is closest to the observer and $\psi=\psi_{\min}=i-\theta$, while at $\varphi=\pi$ the spot is farthest away and $\psi=\psi_{\max}=i+\theta$. 

\begin{figure}
   \centering
   \includegraphics[width=0.4\textwidth]{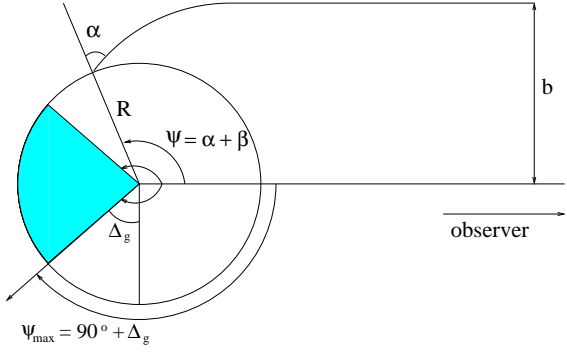}
\caption{Light emitted from a neutron star at an angle $\alpha$ to the normal is observed at impact  parameter $b$, 
with the direction making angle $\psi$ to the spot position vector. The picture is in the plane of photon trajectory.}
         \label{fig:1}
   \end{figure}

Because of gravitational light bending,  photons emitted at an angle  $\alpha$ relative to the local radial direction reach the observer at angle $\psi=\alpha+\beta$ (see Fig. \ref{fig:1}).  The deflection angle $\beta$ reaches the maximum $\Deltag$ when photons are emitted at grazing angles $\alpha=\pi/2$. This defines 
the  visible part of the star: 
\begin{equation}\label{eq_psimax}
\psi < \psi_{\max}=\frac{\pi}{2}+\Deltag, 
\end{equation}
where $\Deltag$ is a function of stellar compactness. 
For a typical neutron star mass of 1.4$\msun$ and radii between 10 and 14 km, the maximum 
bending angle is between 45\degr  and 25\degr.

The relation between $\alpha$ and $\psi$ for slowly rotating stars is given by an elliptical integral \citep{1983ApJ...274..846P}. For stars with radii $R$ larger than about two  Schwarzschild radii $\rg=2GM/c^2$, a very accurate linear relation between the respective cosines can be used instead \citep{2002ApJ...566L..85B,PB06}: 
  \begin{equation}\label{eq_belob}
1-\cos{\alpha}\approx(1-u)(1-\cos{\psi}),
\end{equation} 
where $u=\rg/R$. In this approximation, the visibility condition (\ref{eq_psimax}) gets a simple form
\begin{equation} \label{eq_cospsimax}
\cos{\psi}> \cos{\psi_{\max}}=-\frac{u}{1-u},
\end{equation}
where now 
\begin{equation} \label{eq_darklimit}
\sin{\Deltag}=\frac{u}{1-u}\equiv\kappa.
\end{equation}

\begin{figure}
   \centering
   \includegraphics[width=0.4\textwidth]{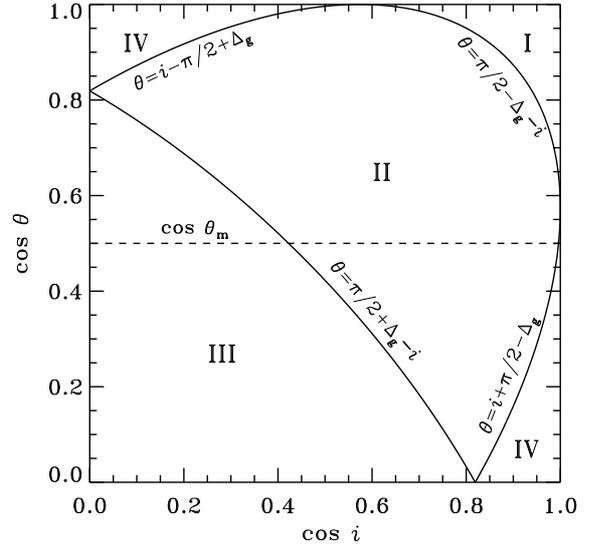}
\caption{Beloborodov's classes of pulsars on the plane ($\cos i,\cos\theta$). 
The visibility classes I--IV  are shown for  a moderate light bending $\Deltag=30^{\circ}$.
If the magnetic inclination is constrained by $\theta<\thetamax$, 
the area below the dashed line $\cos\theta=\cos\thetamax$ is forbidden.}
         \label{fig:2}
   \end{figure}

For the assumed dipole magnetic field with two antipodal spots, a light curve can belong to any of the four visibility classes introduced by \citet{2002ApJ...566L..85B}, according to the values of $i$, $\theta$, and $u$. For a class I pulsar, the secondary pole is invisible and  the primary pole is always visible. Class II pulsars have their primary pole always visible, but the secondary appears and disappears during the rotation period. In class III pulsars both poles appear and disappear 
during the rotational period. Class IV pulsars have both their poles visible at all times. 
See Fig.~\ref{fig:2} for an example of visibility classes for a moderate light bending. 

For random inclination $i$ and magnetic inclination $\theta$, the probability densities $\rmd P/\rmd \cos i$ and  $\rmd P/\rmd \cos \theta$ are constants.  Therefore, the area covered by a certain class on the $(\cos i, \cos\theta)$ plane directly gives the probability of a random pulsar belonging  to that class. Depending on the stellar compactness, the area covered by each class changes, and for example at high $u$,  bending is strong and the area occupied by the class IV pulsars grows. 

For the blackbody emitting spot, the visibility class directly defines the shape of the light curve \citep{2002ApJ...566L..85B}, while for an arbitrary emissivity pattern  the number of light curve classes can differ from the number of visibility classes. 

The observed flux from a small homogeneous spot can be expressed as 
\begin{equation}
F=I \Omega,
\end{equation}
where $I$ is the observed intensity and $\Omega$ is the solid angle covered by the spot on the observer's sky.  
The spot of area $S= R^{2} \rmd \cos\psi \rmd\phi$ seen at impact parameter  $(b,b+db)$ 
occupies a solid angle  $\Omega=b \rmd b \rmd\phi/D^{2}$, where $D$ is the distance to the observer.  
Owing to the gravitational redshift, the bolometric intensity is reduced from the emitted value $I_{0}$ to
\begin{equation}\label{eq_intensity}
I=(1-u)^{2} \ I_{0}(\alpha) ,
\end{equation}
where the emitted intensity can be a function of the emission angle $\alpha$. 
The impact parameter $b$ and the emission angle $\alpha$ are related by 
\begin{equation} \label{eq_alpha}
\sin{\alpha}=\frac{b}{R}\sqrt{1-u}.
\end{equation}
Combining the expressions above we get \citep{2002ApJ...566L..85B,PB06}: 
\begin{equation}\label{dF}
F=(1-u)^{2}I_{0}(\alpha) \frac{1}{1-u} \frac{\rmd\cos\alpha}{\rmd\cos\psi} \ \frac{S\, \cos{\alpha} }{D^{2}}.
\end{equation}
Using approximation~(\ref{eq_belob}), the flux takes the form 
\begin{equation}\label{dFAB}
F=(1-u)^{2}I_{0}(\alpha) \ \frac{S\, \cos{\alpha} }{D^{2}}.
\end{equation}

\subsection{Light curves and probabilities for blackbody spots}

   \begin{figure}
   \centering
   \includegraphics[width=0.4\textwidth]{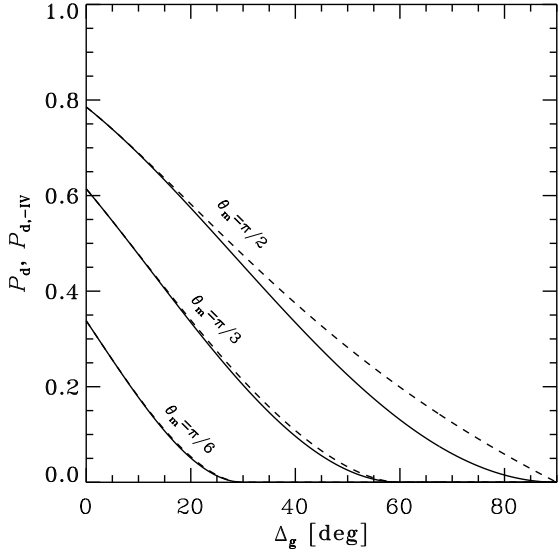}
  \caption{Probability to obtain double-peaked light curves for blackbody spots 
  as a function of the maximal gravitational light deflection angle $\Deltag$.  
  Solid  curves represent the probability given by Eq. (\ref{eq_bbconstr}) when class IV pulsars are included. 
The dashed curves represent the case where class IV pulsars are excluded, given by Eq. (\ref{exclude4a}).
 The curves correspond to three different cases of the maximum magnetic inclination 
 $\thetamax=\pi/2$ (i.e. unconstrained), $\pi/3$ and $\pi/6$.  
 The upper curves are given by analytical relations (\ref{eq_PdmIV}) and (\ref{eq_Pdouble}).}
  \label{fig:3}
   \end{figure}

Let us first assume that the hotspots emit blackbody radiation, i.e. $I_{0}(\alpha)=I_{0}=$const. The flux from a spot is then simply
\begin{equation}
F=(1-u)^{2} \ I_{0}  \frac{S}{D^{2}}  \left[u+(1-u)\cos{\psi}\right] . 
\end{equation}
Below we will use the flux normalized to \(F_{0}\equiv(1-u)^{2}I_{0}S/D^{2}\), i.e 
the flux from the primary spot is 
\begin{equation} \label{eq: belobflux}
F_{\rm p}=u+(1-u)\cos{\psi}= Q+U\cos{\varphi},
\end{equation}
where
\begin{equation}
Q=u+(1-u)\cos{\theta}\cos{i} , \quad 
U=(1-u)\sin{\theta}\sin{i}, 
\end{equation}
and for the secondary spot the flux is 
\begin{equation}
F_{\rm s}=\cos{\alphas}=u-(1-u)\cos{\psi}= 2u-Q-U\cos{\varphi}.
\label{eq_FS}
\end{equation}
Obviously the flux and the variability amplitude depend on the stellar compactness and the position of the spot on the NS surface. As discussed by \citet{2002ApJ...566L..85B}, the light curves from class I pulsars are single-peaked. In class II the light curves are also single-peaked  with a plateau between the pulses. Class III is the only one contributing the double-peaked light curves. Class IV pulsars produce flat light curves, because 
\begin{equation} \label{eq_flat4} 
F_{\rm p}+F_{\rm s}=2u= {\rm const}.
\end{equation} 

In the blackbody case, the probability to observe double-peaked light curves depends only
on the maximal gravitational light-bending angle $\Deltag$. 
Because only class III pulsars produce double-peaked profiles, 
the probability to observe it is given by the area $\Sigma_{\rm{III}}$ occupied by 
class III on $\cos i$--$\cos\theta$ plane (see Fig.~\ref{fig:2}): 
\begin{equation} \label{eq_Pdouble}
P_{\rm{d}}(\Deltag)\! =\!\Sigma_{\rm{III}}\!=\!\!\int\limits_{0}^{\cos{\Deltag}} \!\!\!\!
\cos{\left(\frac{\pi}{2}+\Deltag-i\right)}\,\rm{d}\cos{i}=\left(\frac{\pi}{4}-\frac{\Deltag}{2}\right)\cos{\Deltag},
\end{equation}
which is plotted as a solid curve in Fig.~\ref{fig:3} (the case $\thetamax=\pi/2$). If bending is weak, $\Deltag=0$, the probability is $\pi/4$, i.e. 79\%, and steadily decreases with increasing $\Deltag$. This result differs dramatically from that obtained by B03, who assumed that class II, III, and IV are all producing double-peaked profiles. 

Another way to interpret the model predictions is to exclude the class IV light curves from the light curve analysis since they are flat, i.e. these are not pulsars. It would mean that the total parameter space for the pulsars would be reduced and the fraction of  single- and double-peaked light curves is changed (see Fig.~\ref{fig:2}). The area corresponding to the classes I, II, and III is 
\beq
\Sigma_{\rm{I+II+III}}&=&\int\limits_{0}^{\sin{\Deltag}} \!\! \cos{\left(i-\frac{\pi}{2}+\Deltag\right)} \, 
\rmd \cos {i} +1-\sin{\Deltag} \\
&-&\!\!  \int\limits_{\cos{\Deltag}}^{1} \!\! \cos{\left(i+\frac{\pi}{2}-\Deltag\right)} \, \rmd\cos{i}
=1-\sin{\Deltag}+\Deltag\cos{\Deltag}, \nonumber
\eeq
and the probability to observe a double-peaked light curve is thus
\begin{equation}\label{eq_PdmIV}
P_{\rm{d,-IV}}(\Deltag)=\frac{\Sigma_{\rm{III}}}{\Sigma_{\rm{I+II+III}}}=
\left(\frac{\pi}{4}-\frac{\Deltag}{2}\right)\left[\Deltag+\tan{\left(\frac{\pi}{4}-\frac{\Deltag}{2}\right)}\right]^{-1},
\end{equation}
which is plotted as a function of the maximal gravitational light deflection angle in Fig.~\ref{fig:3} (dashed curve, case $\thetamax=\pi/2$). This probability is only slightly higher than that given by Eq.~(\ref{eq_Pdouble}) where class IV is included  (solid curve).

We need to notice here that class IV pulsars produce flat light curves only if the emission is blackbody-like and if the radiating spots are exactly antipodal. Small deviations from the isotropy, e.g. if we use an Eddington approximation for the intensity, 
$I_0(\alpha)=I_0 (1+ \ani\cos\alpha)$ (where $h$ is a parameter),  cause oscillations even in class IV pulsars. For small $h$, the peak-to-peak amplitude is \citep{PB06}: 
\begin{equation}
A= \frac{F_{\max}- F_{\min} }{F_{\max}+ F_{\min}} = \frac{1-u}{2u} |\ani|  \times 
\left\{ \begin{array}{ll} 
\sin 2i\ \sin 2\theta & \mbox{if $i +\theta < \pi/2$,} \\
\cos^2 (i-\theta)  & \mbox{if $i +\theta > \pi/2$.} 
\end{array}
\right. 
\end{equation}
In addition, even a slight displacement from the exactly antipodal positions leads to pulsation of the light curve.
Let us assume the secondary spot is shifted by $\delta\theta$ and $\delta \varphi$ in latitude and azimuth: 
 $\thetas=\pi-(\theta+\delta\theta)$ and $\phis=\varphi+\pi+\delta \varphi$. It is easy to show that 
 for the blackbody emission 
 \begin{equation}
A= \frac{1-u}{2u} \sin i \ \sqrt{\cos^2\theta \ (\delta\theta)^2 + \sin^2\theta\  (\delta \varphi)^2} . 
 \end{equation}
Thus we see that the class IV pulsars are actually expected to pulsate. They predominantly have single-peaked profiles if 
the radiation emission pattern is close to   black body (or more peaked, see Sect. \ref{sec:cosalpha2} and \ref{sec:beam_cosa_n}), and spots are close to antipodal positions. Equation~(\ref{eq_Pdouble}) should thus give a more realistic probability of observing double-peaked light 
curves.

\begin{figure*}
\centering
\includegraphics[width=0.4\textwidth]{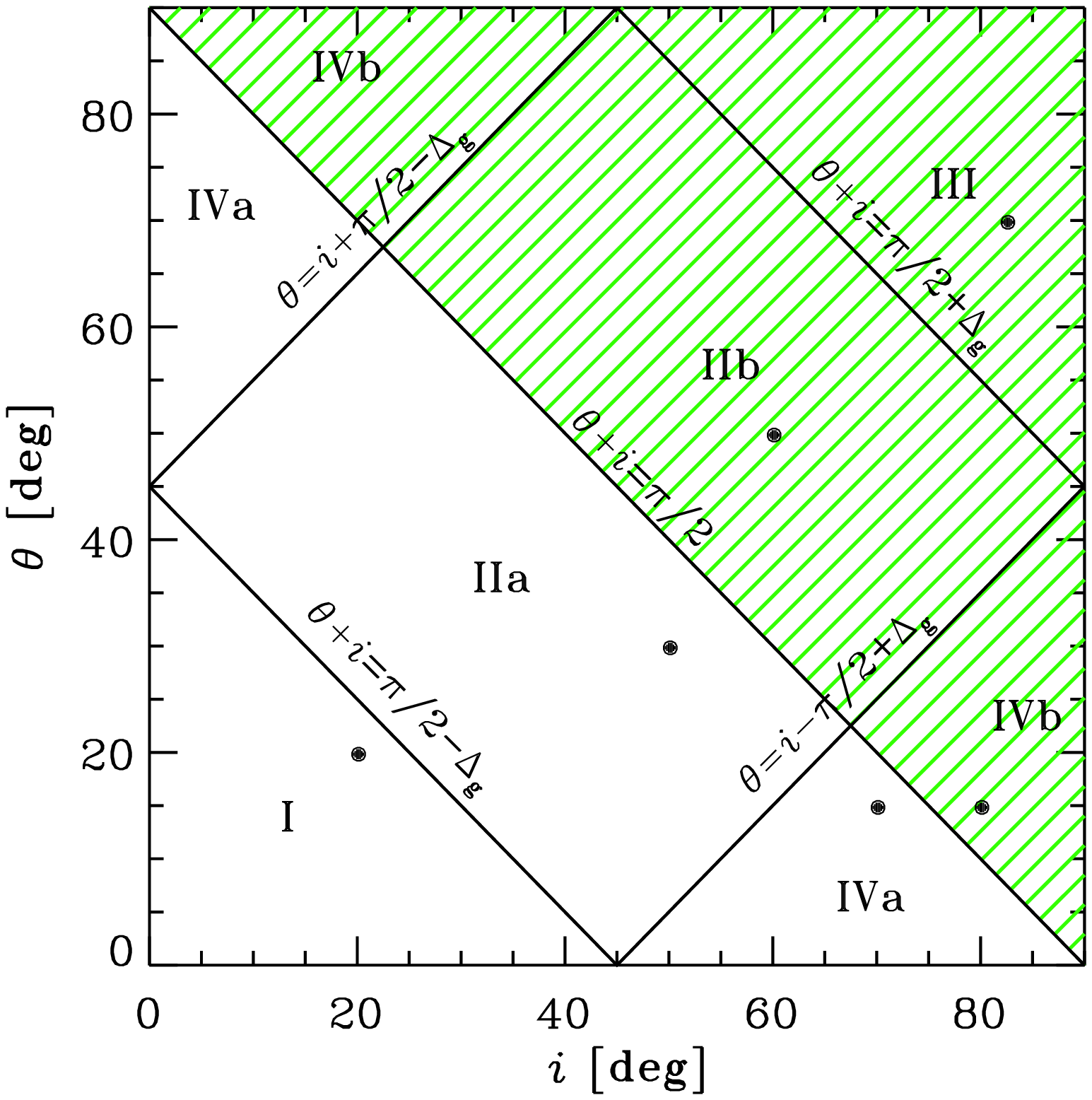} 
\hspace{1cm}
\includegraphics[width=0.4\textwidth]{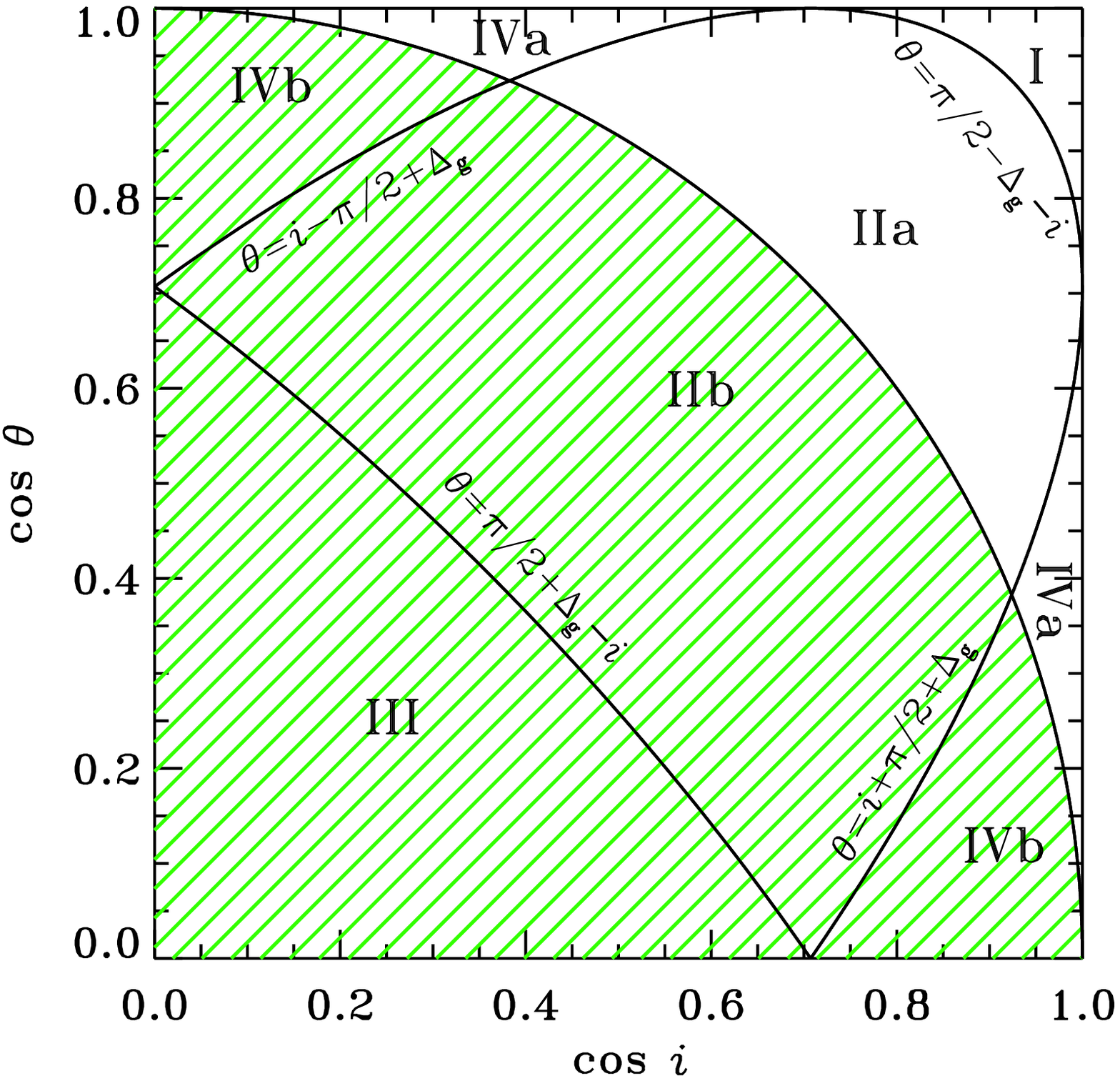}
\caption{Light curve classes for beam pattern $\cos^{2}{\alpha}$ on the $i$--$\theta$ plane (left) and 
the $\cos i$--$\cos\theta$ plane (right). 
The shaded areas correspond to the double-peaked light curves. The
number of light curve classes for this pattern differs from the blackbody case. 
The small filled circles at the left panel correspond to the
parameter pairs ($i$,$\theta$) of the light curves plotted in Fig.~\ref{fig:5}.}
\label{fig:4}
\end{figure*}

\subsubsection{Limiting magnetic inclination}

In the previous discussion, we assumed that magnetic inclination can take any value $\theta\leq\pi/2$. However, as proposed by \citet{1981AAA...102...97W} and B03, there can be an upper limit  to that angle $\thetamax$. Then the fraction of pulsars belonging to different visibility classes is changed (see Fig. \ref{fig:2}).  The constrained $\theta$ case alters the probabilities of observing the double-peaked profiles.

Let us consider a blackbody-like emission pattern and include all the pulsar classes I--IV. The probability to observe double-peaked light curves is proportional to the area of class III above the curve $\cos \theta=\cos \thetamax$ (see Fig. \ref{fig:2}) and is the function of the bending angle $\Deltag$ and  $\thetamax$: 
\beq \label{eq_bbconstr}
P_{\rm{d}}(\Deltag,\thetamax) & = & \frac{\int\limits_{0}^{\cos{\imin}} \cos{\left(\frac{\pi}{2}+\Deltag-i\right)}\,\rmd\cos{i} -\cos{\thetamax} \cos{\imin}}{1-\cos{\thetamax}}  \nonumber \\
&=&\frac{(\thetamax-\Deltag)\cos{\Deltag}+\cos{\thetamax}\sin{(\Deltag-\thetamax)}}{2(1-\cos{\thetamax})},
\eeq
where $\imin=\pi/2 +\Deltag -\thetamax$ is the crossing point between curves 
$\theta=\thetamax$ and $\theta=\pi/2 +\Deltag-i$.  If $\thetamax<\Deltag$, the probability is zero. 

If we exclude class IV from consideration, the corresponding probability becomes 
\begin{equation} \label{exclude4a}
P_{\rm{d},-IV}(\Deltag,\thetamax) = 
\left\{\begin{array}{ll}
 0      &\ \mbox{if}\ \thetamax<\Deltag \\
 P_{1} &\ \mbox{if} \ \Deltag<\thetamax<\frac{\pi}{2}-\Deltag\ \mbox{and}\ \Deltag<\frac{\pi}{4} \\
 P_{2} &\ \mbox{if} \ \frac{\pi}{2}-\Deltag<\thetamax<\frac{\pi}{2}\ \mbox{and}\  \Deltag<\frac{\pi}{4}  \\
 P_{2} &\ \mbox{if}  \ \Deltag<\thetamax<\frac{\pi}{2}\ \mbox{and}\ \Deltag>\frac{\pi}{4},
\end{array}\right.
\end{equation}
where
\begin{equation}
P_{1}=\frac{(\thetamax-\Deltag)\cos{\Deltag}+\cos{\thetamax}\sin{(\Deltag-\thetamax)}}
{2(1-\cos{\thetamax})-(\sin{\Deltag}-\Deltag\cos{\Deltag})}
\end{equation}
and
\begin{equation}
P_{2}=\frac{ (\thetamax-\Deltag)\cos{\Deltag}+\cos{\thetamax}\sin{(\Deltag-\thetamax)} } 
{2(1\!-\!\sin{\Deltag})\!+\!\Big( 2\Deltag\!-\!\frac{\pi}{2}\!+\!\thetamax \!\Big)\cos{\Deltag}\!-\!
\cos{\thetamax}\sin{(\Deltag+\thetamax)}} .
\end{equation}
The probabilities  to observe double-peaked profiles for two cases, including and  excluding the class IV pulsars, 
given by Eqs. (\ref{eq_bbconstr})  and (\ref{exclude4a}),  are shown in Fig.~\ref{fig:3} by solid and dashed curves, respectively. 
The significant difference in these probabilities appears only for a large bending angle $\Deltag$.

   \begin{figure}
   \centering
   \includegraphics[width=0.4\textwidth]{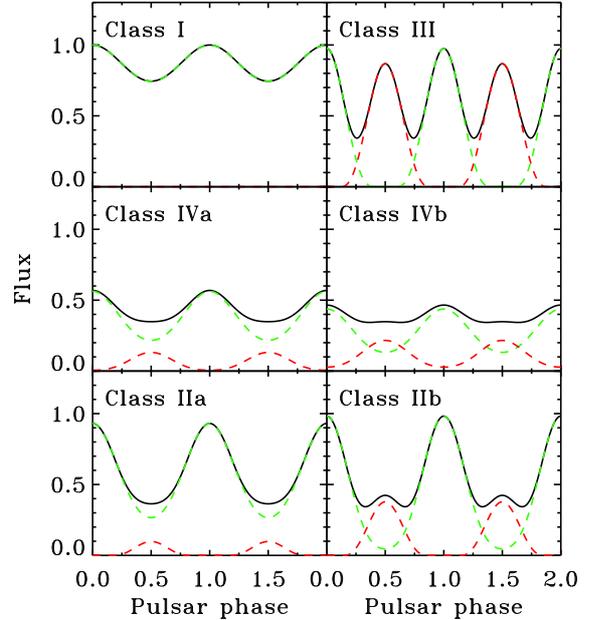}    
  \caption{Light curve classes for beam pattern $\cos^{2}{\alpha}$ in the moderate case of light bending $\Deltag=45^{\circ}$. 
  The light curves are plotted with following parameters: $\theta=20^{\circ}$ and $i=20^{\circ}$ (class I), $\theta=70^{\circ}$ and $i=82^{\circ}$ (class III),   $\theta=15^{\circ}$ and $i=70^{\circ}$ (class IVa),  $\theta=15^{\circ}$ and $i=80^{\circ}$ (class IVb),  $\theta=30^{\circ}$ and $i=50^{\circ}$ (class IIa), 
  $\theta=50^{\circ}$ and $i=60^{\circ}$ (class IIb).}
  \label{fig:5}
   \end{figure}

\subsection{Modified pencil beam pattern}
\label{sec:modif}

\subsubsection{Beam pattern $\cos^{2}{\alpha}$}
\label{sec:cosalpha2}

Let us now consider two point-like antipodal spots with the emitted intensity given by $I_{0}(\alpha)=I_0\cos{\alpha}$. The light curves produced by the two spots with this more radially concentrated beam-pattern $\cos^{2}{\alpha}$ can be divided into six different classes based on the visibility of the spots and whether the light curves are single- or double-peaked (see Figs.~\ref{fig:4} and~\ref{fig:5}). 
The normalized flux from the primary spot  in  Beloborodov's approximation (\ref{eq_belob}) is  
\begin{equation}
F_{\rm p}=\cos^{2}{\alpha}=(Q+U\cos{\varphi})^{2} .
\end{equation}
The flux  from the secondary pole is 
\begin{equation}
F_{\rm s}=\cos^{2}{\alphas}=(2u-Q-U\cos{\varphi})^{2}.
\end{equation}
In class I, only the primary spot is visible and the light curve is single-peaked with the maximum at pulsar phase $\varphi=0$ and the minimum at $\varphi=\pi$. In class IV, both spots are visible all the time and the flux $F_{\rm p}+F_{\rm s}$ 
has local extrema when
\begin{equation}
\frac{{\rmd}(F_{\rm p}+F_{\rm s})}{\rmd\varphi}=4U\sin{\varphi} \ (u-Q-U\cos{\varphi})=0 ,
\end{equation}
i.e. at $\varphi=0,\pi$, or
\begin{equation} \label{eq:cosphi_uQU}
\cos{\varphi}=\cos{\varphi_{\min}}\equiv \frac{u-Q}{U} = \cot i\  \cot \theta. 
\end{equation}
The latter extrema exist only when 
\begin{equation} \label{eq:ithetapi2}
i+\theta>\pi/2
\end{equation}
and then the light curve is double-peaked. Similarly for class II pulsars, for which the primary spot is always visible and the secondary spot is seen at phases around $\varphi\sim\pi$, we get that the light curve is single-peaked when $i+\theta<\pi/2$ and double-peaked otherwise. 
Condition (\ref{eq:ithetapi2}) divides both classes II and IV into two subclasses according to the number of peaks: 
in IIa and IVa profiles are single-peaked, while in  classes IIb and IVb, they 
are double-peaked. 
In class III, both spots appear and disappear from the view at some pulsar phases, and it is easy to show that both
$\varphi=0$ and $\varphi=\pi$ are maxima, and the light curve is always double-peaked.

Thus the  light curve in classes I, IIa, and IVa  is single-peaked and the profile in III, IIb, and IVb  is double-peaked. The probability to observe double-peaked light curves is now independent of the neutron star compactness  (see  the hatched area in Fig.\ref{fig:4}): 
\begin{equation}
P_{\rm{d}}=\int^{1}_{0} \cos{\left( \frac{\pi}{2}-i\right)} \rmd \cos {i}=\frac{\pi}{4}\approx0.79. 
\label{pdoubleconst}
\end{equation}
This coincides with expression (\ref{eq_Pdouble}) for the Newtonian case, $\Deltag=0$, and the blackbody radiation pattern. 

\begin{figure}
   \centering
   \includegraphics[width=0.4\textwidth]{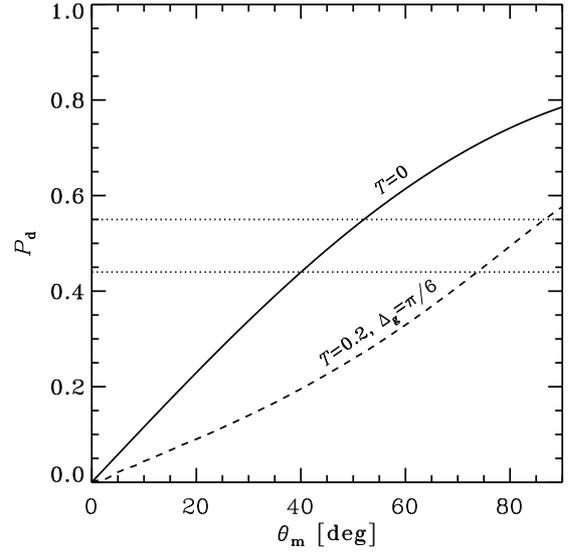}    
\caption{Probability to observe double-peaked light curves as a function of the maximum magnetic inclination $\theta_{\rm{m}}$. 
The solid curve corresponds to any  modified pencil beam patterns discussed in Sect. \ref{sec:modif}, $\cos^{n}{\alpha}$ and $\cos{\alpha}(1+h\cos{\alpha})$ (Eq.~\ref{pcos2}) in the case of full detectability of the secondary pulse, $T=0$. 
The dashed curve shows an example of the beam pattern $\cos^{2}{\alpha}$ with non-zero threshold $T=0.2$ and a moderate bending angle $\Delta_{g}=30^{\circ}$.
  The  dotted lines correspond to the observed probability of having double-peaked
 profiles as given by the classification of data $p_0=0.44$  and $p_1=0.55$ (see Sect. \ref{sec:data}).  
 }
  \label{fig:6}
   \end{figure}

If the magnetic inclination $\theta$ is constrained to lie in the interval $(0,\thetamax)$,  the probability to observe a double-peaked light curve is (shown by the solid curve in Fig.~\ref{fig:6}) 
\begin{equation} \label{pcos2}
P_{\rm{d}} (\thetamax) =\frac{1}{4}\frac{2\thetamax-\sin{2\thetamax}}{1-\cos{\thetamax}}.
\end{equation}
Decreasing the value of $\thetamax$ reduces the probability to obtain double-peaked light curves and for $\thetamax\ll 1$,  $P_{\rm{d}} (\thetamax) \approx \frac{2}{3} \thetamax$. 

\subsubsection{Beam pattern $\cos^{n}{\alpha}$}
\label{sec:beam_cosa_n}

Let us assume that the emission pattern can be described as $I_{0}(\alpha)=I_0\cos^{n-1}{\alpha}$, where $n\ge 1$. The observed flux is then $F\propto I_{0}(\alpha)\cos{\alpha}= I_{0}\cos^{n}{\alpha}$.  The normalized flux from the primary pole is then 
\begin{equation}
F_{\rm p}=\cos^{n}{\alpha}=(Q+U\cos{\varphi})^n
\end{equation}
and the secondary pole has the flux
\begin{equation}
F_{\rm s}=\cos^{n}{\alphas}=(2u-Q-U\cos{\varphi})^n.
\end{equation}
Once the power law index $n$ exceeds 1, the light curves become similar to the case of $n=2$, i.e. the $\cos^{2}{\alpha}$ beam pattern. 
Double-peaked profiles are obtained in the region $i+\theta>\pi/2$.  
Therefore the probability to observe double-peaked light curves in the case of $n>1$ and randomly distributed angles $i$ and $\theta$ 
is $P_{\rm{d}}=\pi/4$.  
For the constrained magnetic dipole inclination,  this probability  is given by Eq.~(\ref{pcos2}).

\subsubsection{Eddington approximation  $\cos{\alpha}(1+h \cos{\alpha})$}

So far we have considered modified pencil beam patterns with varying power index $n$. Let us now assume the radiation intensity deviates from blackbody according to the Eddington approximation $I_{0}(\alpha)= I_{0}(1+h\cos{\alpha})$ and the observed flux $F\propto I_{0}(\alpha)\cos{\alpha}= I_{0}\cos{\alpha}(1+h\cos{\alpha})$. The normalized primary pole flux is 
\begin{equation}
F_{P}=(Q+U\cos{\varphi})[1+h(Q+U\cos{\varphi})]
\end{equation}
and the secondary pole gives
\begin{equation}
F_{S}=(2u-Q-U\cos{\varphi})[1+h(2u-Q-U\cos{\varphi})].
\end{equation}

The shape of the light curves is determined by the anisotropy parameter $h$. 
The observations show that $h$ should be positive rather than negative, because the
negative values can produce double-, triple- and quadruple-peaked light
curves, which are not consistent with the observations. Even 
if there are only single- and double-peaked light curves (for
example when $h=-1$), these double-peaked light curves are double-horned
with equal primary and secondary peak amplitudes. This is not what
we generally observe.

Therefore, we consider positive values  of $h$. The light curves then resemble the classes obtained for modified beam pattern $\cos^{n}{\alpha}$. Double-peaked light curves are only obtained when $i+\theta>\pi/2$ and single-peaked curves when $i+\theta<\pi/2$. Again, this result does not depend on the compactness. Therefore, if angles $\theta$ and $i$ can vary randomly between zero and $\pi/2$, we would observe about $79\%$ double-peaked light curves as calculated earlier (see Eq.~\ref{pdoubleconst}). If the angle between the magnetic field and rotational axis is constrained, then the probability of having double-peaked light curves is just a function of $\thetamax$ as found earlier (see Eq.~\ref{pcos2}). As can be seen, the probabilities to observe double-peaked light curves do not depend on the constant $h$ when it is positive.

\begin{figure*}
\centering
\includegraphics[width=0.4\textwidth]{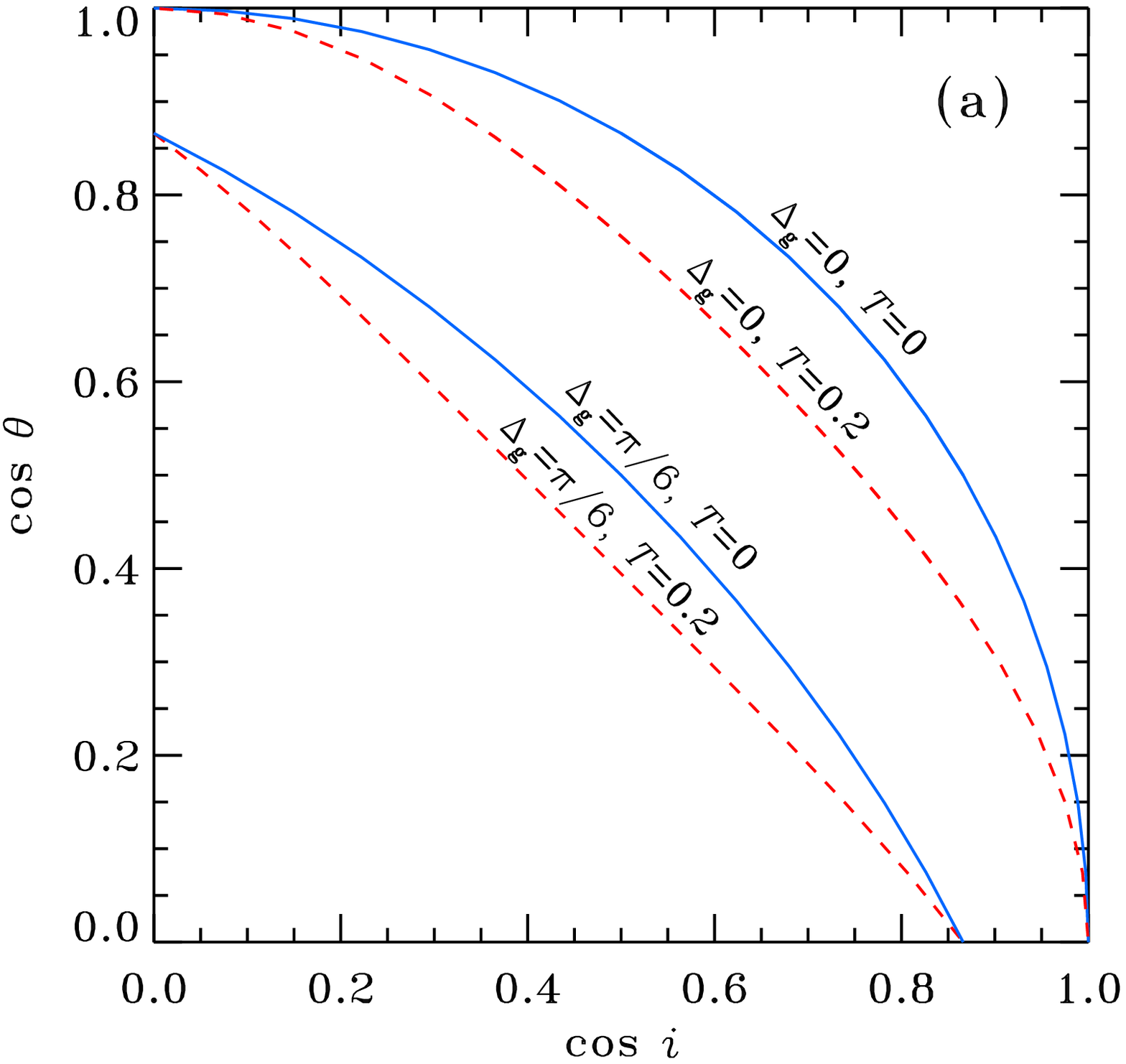} 
\hspace{1cm}
\includegraphics[width=0.4\textwidth]{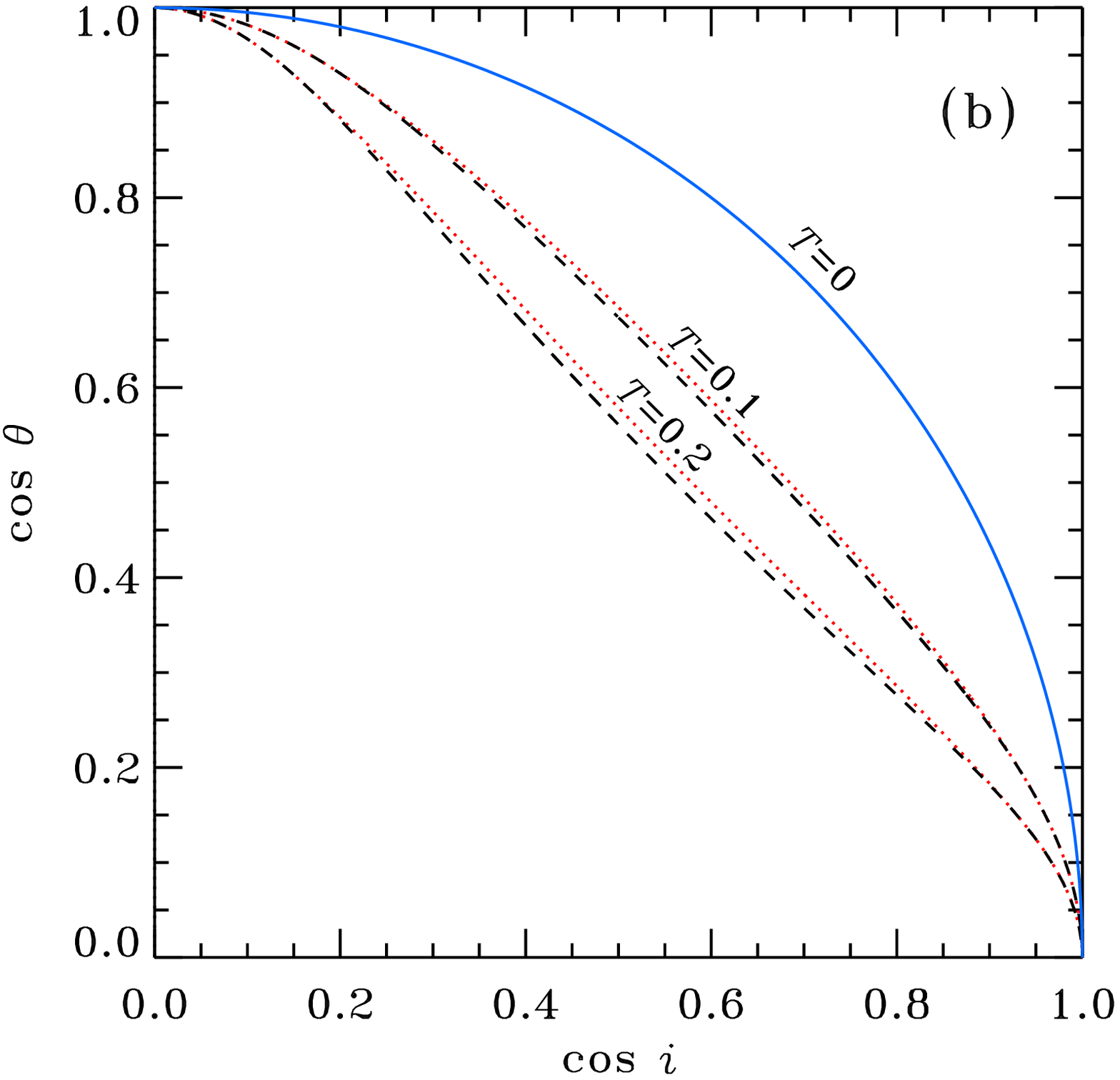}
\caption{Effect of the detection threshold on the fraction of the double-peaked profiles. 
(a)  The parameter space at the $\cos i$--$\cos\theta$ plane for the  blackbody  emission pattern. 
 Upper curves are for the Newtonian case $\Deltag=0$ and the lower curves are 
for the moderate light bending $\Deltag=30^{\circ}$. 
 The solid curves correspond to the zero threshold and the dashed curves are for $T=0.2$. 
 The areas below respective curves correspond to the probability of observing
 double-peaked profiles. 
(b) Same as (a), but for the pencil-beam $\cos^2 \alpha$.  
For the zero threshold the curve is the same for any neutron star compactness. 
The curves for thresholds $T$= 0.1 and 0.2 are shown by  dotted and dashed curves, respectively.
The dotted curves are for $\Deltag=30\degr$, while the dashed ones are for $\Deltag=45\degr$. 
Even then the dependence on compactness is very weak.  
}
\label{fig:7}
\end{figure*}

\subsection{Effect of the detection threshold}

Limited photon statistics and (red-noise) flux variability  can affect the visual assignment of pulsars into single- and double-peaked  classes. 
If  the secondary pulse is hardly visible above the constant emission level, the pulsar is likely to be classified as single-peaked. It is difficult to quantify these effects. Following B03, we can assume that the secondary pulse is visible if its strength above the minimum flux is at least fraction $T$ of  the strength of the primary maximum.  
The probability to observe double-peaked profiles decreases with increasing detection threshold $T$.

\subsubsection{Detection threshold in the blackbody case}

In the blackbody case, for the zero threshold $T=0$ the region, where profiles are double-peaked, coincides with class III. 
There the primary maximum is reached at phase $\varphi=0$,  and its amplitude above the constant flux level is
 $F_{\rm p}(\varphi=0)-2u$ (see Eq.~\ref{eq_flat4}).  For the secondary maximum to  be detected, its amplitude, $F_{\rm s}(\varphi=\pi)-2u$,  should exceed the threshold $T\times [F_{\rm p}(\varphi=0)-2u]$. 
Thus for the non-zero $T$, the region of double-peaked profiles is reduced as the following condition must be satisfied 
\begin{equation}\label{eq_t1}
\frac{F_{\rm s}(\varphi=\pi)-2u}{F_{\rm p}(\varphi=0)-2u}=\frac{\kappa +\cos{(i+\theta)}}{\kappa- \cos{(i-\theta)}}>T.
\end{equation}
Generally  this inequality must be solved numerically. The resulting constraints are shown in Fig.~\ref{fig:7}a. 

In the Newtonian limit $\Deltag=0$ (i.e $\kappa=0$), this inequality is reduced to  
\begin{equation}\label{eq:suhde}
\tan{i}\tan{\theta}> \frac{1+T}{1-T} .
\end{equation}
Then the probability to observe double-peaked light curves for randomly distributed $i$ and $\theta$ is given 
by the integral 
\beq\label{S1}
P_{\rm{d}}&=&\int_0^1 \cos{\theta}\,\rmd\cos{i}=
\int_{0}^{1} \frac{\rmd \cos{i}}{\sqrt{1+\left(\frac{1+T}{1-T}\right)^2\cot^2{i} }} 
\nonumber \\
&=& \frac{1-T^2}{4T} \left[ F(k)- E(k) \right] , 
\eeq
where $k=\sqrt{4T}/(1+T)$, $F(k)$ and $E(k)$ are the complete elliptic integrals. 

Figure~\ref{fig:7}a demonstrates how the parameter space is reduced  where profiles are double-peaked when the detection threshold is increased from zero to $T=0.2$ in a Newtonian- ($\Deltag=0$) and  in a moderately relativistic case, $\Deltag=30^{\circ}$. The solid curves correspond to the total detectability $T=0$ and the areas below them correspond to the probabilities to observe double-peaked light curves. The dashed curves correspond to the threshold $T=0.2$.

\subsubsection{Detection threshold for $\cos^{n}{\alpha}$ beam pattern}
\label{sec:n_det}

Let us consider  the beam pattern $\cos^{n}{\alpha}$, where the index $n$ is greater than 1. We showed in Sect. \ref{sec:beam_cosa_n} that the resulting profile becomes double-peaked when $i+\theta>\pi/2$. The effect of the detection threshold on the visibility of the secondary pulse has to be considered in three different light curve classes {II}b, {III}, and {IV}b (see Fig.~\ref{fig:4}). 

For a blackbody pattern when both poles are visible the flux is constant and equals $2u$. For $n>1$, the minimum flux of $2u^n$ is reached at the 
phase $\varphi_{\min}$ given by Eq. (\ref{eq:cosphi_uQU}) in all three considered classes (because the secondary pole is visible at this phase).

   \begin{figure*}
   \centering
   \includegraphics[width=0.4\textwidth]{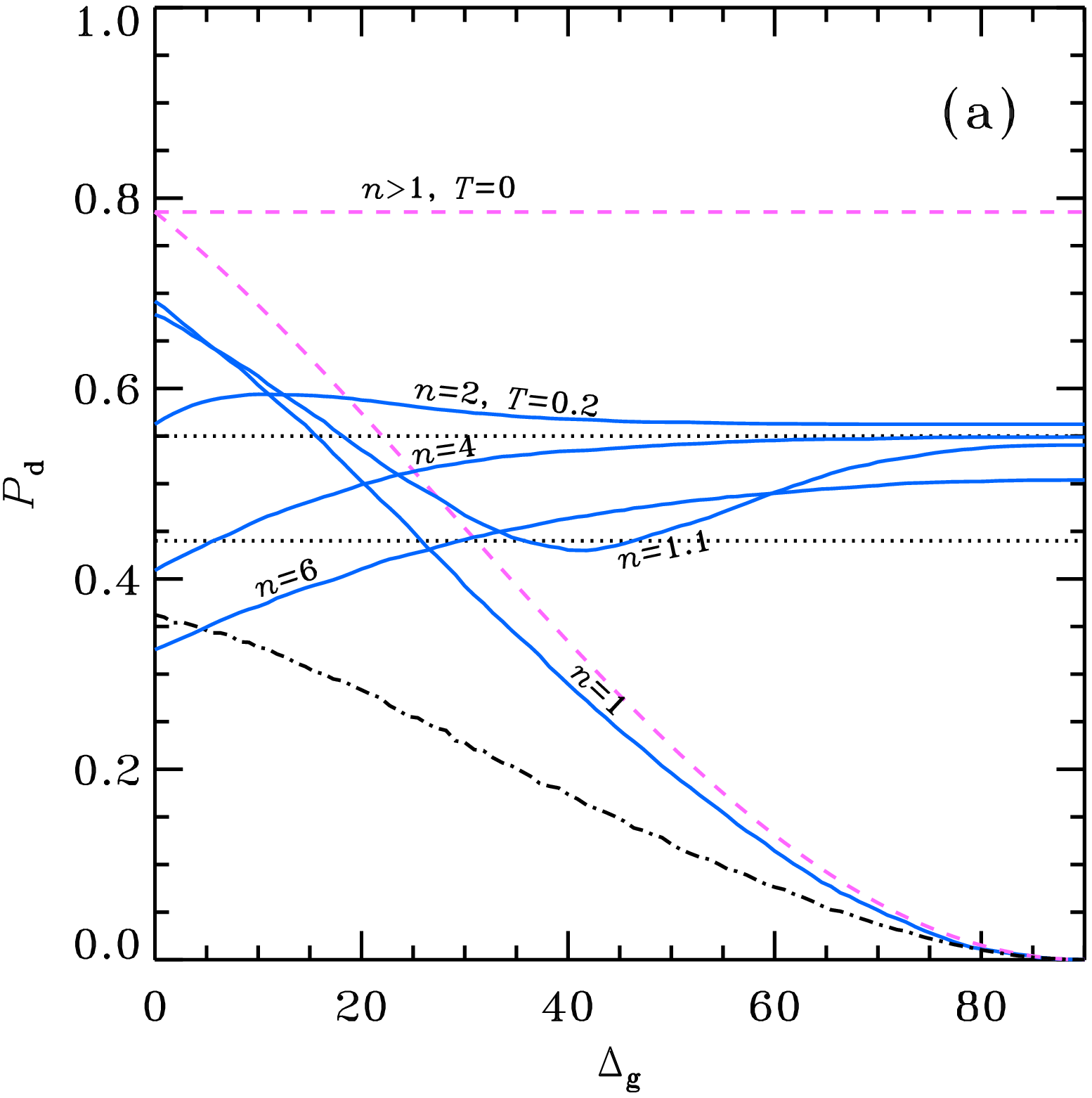}
   \hspace{1cm}
      \includegraphics[width=0.4\textwidth]{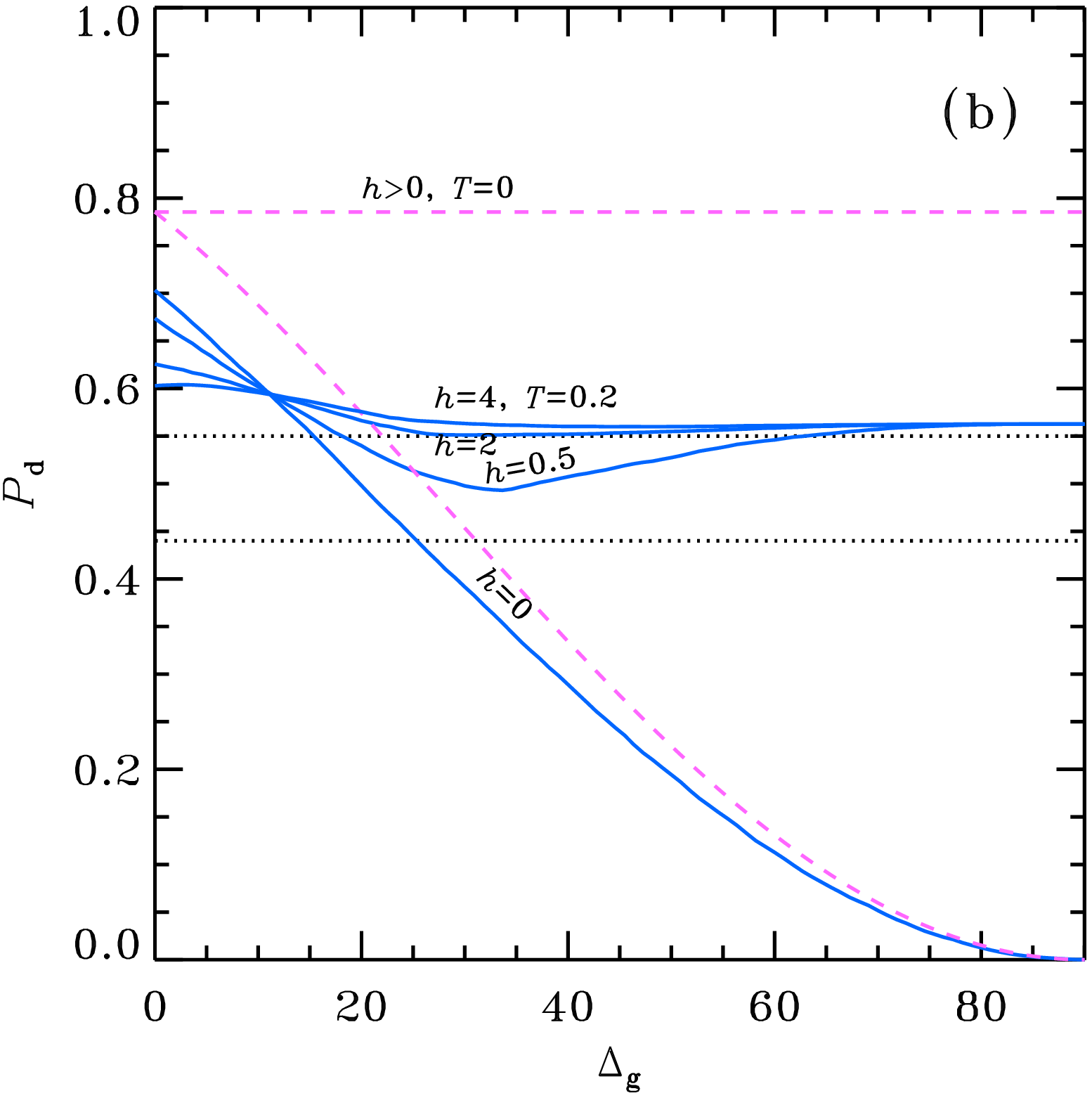}
  \caption{(a) Probability to observe double-peaked profiles for the blackbody emission pattern ($n=1$) as well as for 
  beam patterns with indices $n$=1.1, 2, 4 and 6 for two different threshold cases $T=0$ and 0.2. 
  The dashed curves correspond to $T=0$, while the solid curves are for $T=0.2$.
  The  dotted lines correspond to the observed probability of having double-peaked
 profiles as given by the classification of data $p_0=0.44$ and $p_1=0.55$  (see Sect. \ref{sec:data}).  
The dash-dotted line correspond to the probability to have multiple-peaked light curves for  two 
randomly positioned spots radiating as black bodies. \protect\\
(b) Same as panel (a), but for beam patterns in the Eddington approximation.
 Cases  with anisotropy parameter $h$=0, 0.5, 2 and 4 for two different thresholds $T=0$ and 0.2 are shown.}
  \label{fig:8}
   \end{figure*}

For class {II}b, the primary maximum is reached at $\varphi=0$, when only the primary pole is visible. 
On the other hand, both poles contribute to the flux at the secondary maximum $\varphi=\pi$.   Thus the secondary pulse will be detected if 
\beq \label{eq:IIb_ntr}
&&\frac{F_{\rm p}(\varphi=\pi)+F_{\rm s}(\varphi=\pi)-2u^n}{F_{\rm p}(\varphi=0)-2u^n}  \nonumber \\
 &=& \frac{(Q-U)^{n}+(2u-Q+U)^{n}-2u^{n}}{(Q+U)^{n}-2u^n}> T.
\eeq 
For class III,  the primary maximum is produced by  the primary pole and the secondary maximum by the secondary pole only, thus the condition for the secondary pulse to be seen is
\begin{equation} \label{eq:III_ntr}
\frac{F_{\rm s}(\varphi=\pi)-2u^n}{F_{\rm p}(\varphi=0)-2u^n}=\frac{(2u-Q+U)^n -2u^n}{(Q+U)^n -2u^n}> T.
\end{equation}
Class IV pulsars have both their primary and secondary poles contributing to the flux all the time, and therefore the threshold to the secondary pulse to be detected is
\beq\label{eq:IV_ntr}
&&\frac{F_{\rm p}(\varphi=\pi)+F_{\rm s}(\varphi=\pi)-2u^n}{F_{\rm p}(\varphi=0)+F_{\rm s}(\varphi=0)-2u^n} \nonumber \\
&=& \frac{(Q-U)^{n}+(2u-Q+U)^{n}-2u^{n}}{(Q+U)^{n}+(2u-Q-U)^n -2u^n}> T.
\eeq

For $n=2$ the boundary between single- and double-peaked profiles given by relations (\ref{eq:IIb_ntr})--(\ref{eq:IV_ntr})  are shown in  Fig.~\ref{fig:7}b at the $\cos i$--$\cos\theta$ plane. The areas  below the corresponding curves give the probability $P_{\rm d}$ to observe double-peaked profiles.  Figure~\ref{fig:8}a shows this probability as a function of compactness for various indices $n$ and two  thresholds $T=0$ and 0.2. When $T=0$, i.e. even the smallest bumps in the light curve can be detected, all beam patterns with $n>1$ predict  $P_{\rm d}= 79\%$. 
For $T=0.2$, the probability to observe double-peaked profiles decreases. 
For typical $\Deltag\in[25\degr,45\degr]$, a slight deviation from $n=1$ to $n=1.1$ immediately increases the value of $P_{\rm d}$
from $\sim$0.3 to 0.45. The maximum of $P_{\rm d}\sim0.6$ is reached at $n\sim 2$, and at $n>2$ the probability starts decreasing again.
This probability  depends  weakly on $\Deltag$  as well as on index $n$ (see Fig.~\ref{fig:7}b and Fig.~\ref{fig:8}a). 
If the magnetic inclination is constrained, the fraction of double-peaked profiles decreases as shown in Fig. \ref{fig:6} for the $n=2$ case.

\subsubsection{Detection threshold in the Eddington approximation case}
\label{sec:Edd_det}

The emission pattern  in the form given by the Eddington approximation produces 
light curve classes identical to the classes for the modified pencil beam $\cos^{n}{\alpha}$. Therefore the consideration of the threshold is similar to the previous section. Double-peaked light curves are obtained only in classes II{b}, III, and IV{b}, and   increasing the threshold will increase the probability of observing single-peaked light curves in these classes.

The local extrema are at the same pulsar phases as for the $\cos^{2}{\alpha}$ beam pattern, i.e at $\varphi=0,\pi$ and $\cos{\varphi}=(u-Q)/U$. At the minima the flux is 
\begin{equation}
F_{P}\left(\cos{\varphi}=\frac{u-Q}{U}\right)+F_{S}\left(\cos{\varphi}=\frac{u-Q}{U}\right)=2u(1+hu).
\end{equation}   

In class IVb, the secondary pulse is detected if the following condition holds:
\beq
\frac{F_{\rm p}(\varphi=\pi)+F_{\rm s}(\varphi=\pi)-2u(1+hu)}{F_{\rm p}(\varphi=0)+F_{\rm s}(\varphi=0)-2u(1+hu)} 
= \frac{\cos^{2}{(i+\theta)}}{\cos^{2}{(i-\theta)}}> T.
\eeq
In class IIb, the secondary pulse is detected if
\beq 
&&\frac{F_{\rm p}(\varphi=\pi)+F_{\rm s}(\varphi=\pi)-2u(1+hu)}{F_{\rm p}(\varphi=0)-2u(1+hu)}  \nonumber \\
 &=& \frac{h[(Q-U)^{2}+(2u-Q+U)^{2} -2u^{2}]}{Q+U-2u(1+hu)+h(Q+U)^{2}}> T.
\eeq 
Finally in class III, the secondary pulse is detected if
 \beq 
&&\frac{F_{\rm s}(\varphi=\pi)-2u(1+hu)}{F_{\rm p}(\varphi=0)-2u(1+hu)}\nonumber\\
&=&\frac{-Q+U+h(2u-Q+U)^{2}-2hu^{2}}{Q+U+h(Q+U)^{2}-2u(1+hu)}> T.
\eeq

The influence of the increasing threshold on the probability to observe
 double-peaked light curves for various anisotropy parameters $h$ is shown in Fig.~\ref{fig:8}b. 
 The case $h=0$ corresponds to the blackbody emission pattern and therefore gives the same dependences as
 plotted in Fig.~\ref{fig:8}a (curves marked with $n=1$).
With full detectability of the secondary pulse $T=0$ the probability to observe double-peaked 
 profiles is constant  $P_{\rm{d}}\approx79\%$  for any $h>0$.  
 This is similar to the result obtained for the modified pencil beam pattern $\cos^{n}{\alpha}$ with index $n>1$  
 discussed in Sect. \ref{sec:n_det}. 

When the threshold $T$ is increased, the probability to have
 double-peaked light curves falls and becomes almost independent of the bending angle $\Deltag$ and anisotropy parameter $h$.

\section{Results}
\label{sec:results}

Based on our classification of pulsars into single- and double-peaked in Sect. \ref{sec:data}, we can now compare the corresponding fractions with the predictions of various radiation models.

\subsection{Blackbody emission pattern and comparison to B03}

\begin{figure}
   \centering
   \includegraphics[width=0.4\textwidth]{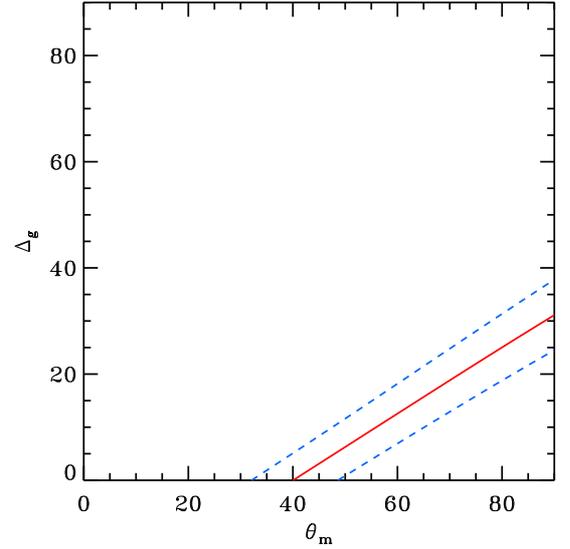}
  \caption{Constraints on the maximum magnetic inclination $\thetamax$ and maximum deflection angle $\Deltag$ 
 for the blackbody pattern and full detectability of the secondary pulse. The solid line gives the most probable values and
 the dashed lines bound  the 95\% confidence region.  }
   \label{fig:9}
   \end{figure}

\begin{figure}
   \centering
   \includegraphics[width=0.4\textwidth]{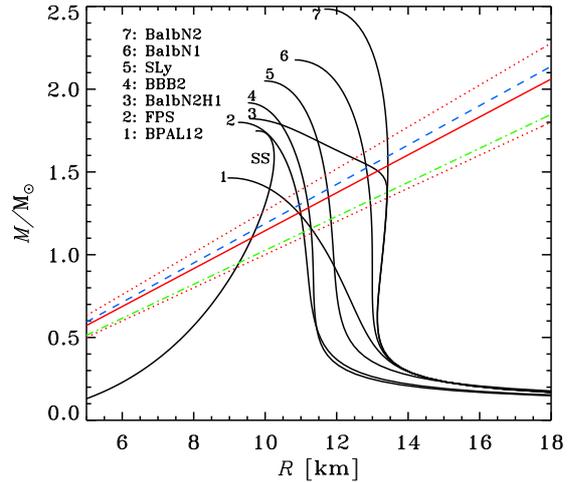}
  \caption{Different mass-radius relations of neutron and strange stars equation of states (curves 1--7 and SS, see  \citealt{2006MNRAS.369.2036S})  and the obtained constraints on the compactness assuming a blackbody emission pattern and unconstrained 
  magnetic inclination. 
  The most probable value for the compactness ($\Deltag=31\degr$) is shown by the solid red line and the 95\% confidence region is bound by   red dotted lines.  The dot-dashed-line shows the most probable value for the compactness (corresponding to $\Deltag=26^{\circ}$) obtained for the detection 
  threshold $T=0.2$. The blue dashed line corresponds to the upper limits on mass as a function of radius obtained by B03.  }
  \label{fig:10}
   \end{figure}

Let us consider first the blackbody radiation pattern and assume that secondary pulses are fully detectable (i.e. $T=0$). If the magnetic inclination is unconstrained (i.e. $\thetamax=\pi/2$), taking $p=p_0$ (Eq.\ref{eq:p0_obs}) we get the maximum deflection angle 
$\Deltag=31\degr\pm3\degr$ (1$\sigma$ error), if we include class IV pulsars into consideration, or $\Deltag=33\degr\pm4\degr$ if we do not, 
see upper curves in Fig. \ref{fig:3}. As the effect of excluding class IV pulsars is small, we take below that class into consideration. 
If we assume that there is an upper limit of magnetic inclination $\thetamax$, we get that $\thetamax>32\degr$ (at 95\% confidence) 
for any deflection angle, and for the Newtonian case (i.e. $\Deltag=0$) we have $\thetamax=40\degr\pm4\degr$. In general, the probability to observe a double-peaked profile was found to be given by Eq.~(\ref{eq_bbconstr}), and the corresponding constraints on both $\thetamax$ and $\Deltag$ are shown Fig.~\ref{fig:9}. 
For typical neutron stars with $\Deltag\in[25\degr,45 \degr]$, we have $\thetamax>70\degr$ (at 95\% confidence).  
Thus our results favor not very compact neutron stars (with radius of above 12 km for $M=1.4\msun$) and nearly random magnetic dipole inclinations. 
Figure~\ref{fig:10} shows various  neutron star mass-radius relations and our best constraint corresponding to  $\Deltag=31\degr$.

Let us now look at the effect of the detection threshold which reduces 
the probability to observe double-peaked profiles (see Fig.~\ref{fig:8}a). For unconstrained magnetic inclination, we evaluate the probability to see double-peaked profiles by numerically computing the area on the $\cos i$--$\cos\theta$ plane where inequality (\ref{eq_t1}) is satisfied. Compared to the case $T=0$, the maximum bending angle  for $T=0.2$ decreases  by just a few degrees to $\Deltag=26\degr\pm3\degr $, which (for fixed $M$) would favor even larger neutron star radii (see Fig. \ref{fig:10}).

A similar study was made by B03, who obtained a different result for a number of reasons. 
Although our classification of X-ray pulsar light curves differs from B03 (we have 60 double-peaked profiles out of 124, while B03 have 38/88),
the probabilities $p_0$ are similar within a few percent.  However, B03 assumed that single-peaked light curves would be observed only if one magnetic pole is seen (class I). This led to the conclusion that the probability of having single-peaked profiles is about 21\%  for $\Deltag=0$ and 
this  value decreases as compactness grows. Therefore in their point of view there is a conflict between the theory and observations, because the percentage of observed single-peaked light curves is much higher. As a result, B03 needed to constrain the magnetic field geometry (by introducing maximum magnetic inclination) in order to attain consistency between the observations and the theory.
In our view, class III is the only place to obtain double-peaked light curves. Therefore, we get $\thetamax\gtrsim35\degr$ (and for reasonable value of the compactness the magnetic dipole inclination is random),  while B03 got  $\thetamax <50\degr$ (compare our Fig. \ref{fig:9} to Fig. 7 in B03).

 \subsection{Modified pencil-beams and implications for neutron star compactness
 and magnetic field geometry}

The pulsars do not shine as blackbodies and therefore we now take a look at the effect of changing the emission pattern. For  the $\cos^{2}{\alpha}$ beam pattern, the probability of having double-peaked profiles is given by Eq.~(\ref{pcos2}). If magnetic inclination is not constrained, $P_{\rm d}=0.79$, which significantly exceeds the observed fraction. Varying $\thetamax$, we can get an agreement between the observations and the model. 
Requiring $P_{\rm d}=p_0$ (see Fig. \ref{fig:6}), we get $\thetamax=40\degr\pm4\degr$, while if we account only for 
pulsars observed at energies above 10 keV and take $P_{\rm d}=p_1$, then $\thetamax=51\degr\pm4\degr$. 
We note here that the probability to observe  double-peaked profiles (\ref{pcos2}) does not depend on the deflection angle $\Deltag$ and thus we cannot get any constraints on the compactness of the neutron star (for a full detectability $T=0$). 
Other pencil beam models, $\cos^{n}{\alpha}$ and $\cos{\alpha}\,(1+h\cos{\alpha})$, give identical results. 
Thus even if the parameters describing the beam patterns of various NS have a large spread, the results are not affected.

The detection threshold may play an important role in classifying the pulse profiles.  
The expectation value for the fraction of the double-peaked profiles becomes lower and makes it rather 
close to the observed one for $T=0.2$, even if deviations from the blackbody pattern are not very large  (see Fig. \ref{fig:8}). 
Taking $n=2$, the predicted fraction of double-peaked profiles is about 0.57, which is within 3$\sigma$ of the observed value $p_0=0.44$ and within 1$\sigma$ of the value $p_1=0.55$ obtained for pulsars observed above 10 keV (see Table \ref{pulsartable1}). 
Increasing $n$ makes the agreement even better. 
Again we stress that the predicted fraction depends only weakly on the neutron star compactness, which therefore cannot be constrained. 
Another conclusion is that the magnetic dipole inclination can be random. 
These results are not affected by our assumptions about the emission pattern if it is sufficiently far from the blackbody, 
because indices $n$ in the interval between about 1.5 and 6 and the anisotropy parameter $h>0.5$ predict similar fractions of 
double-peaked profiles  (see Fig. \ref{fig:8}).

The observed light curves are not always symmetric indicating the secondary pole is not at the antipodal position.
To check how our assumption on antipodal positions of the spots affect the results, 
we have simulated the pulse profiles for two randomly positioned spots on the neutron star surface radiating as blackbodies. 
The probability to observe multiple-peaked light curves depends on the compactness of the neutron star reaching the maximum of 37\% in the Newtonian limit (see the dashed-dotted curve in Fig.~\ref{fig:8}a). This is much less than what is observed (between 44\% and 55\%). 
A significant predicted fraction of multiple-peaked profiles (with number of peaks three or larger) even for the blackbody pattern 
contradicts the fact that such profiles are not observed. Other, more beamed patterns will just increase the fraction of multiple-peaked profiles. 
Thus even though the observed light curves indicate slight asymmetry, the simulations show that the spot positions cannot be completely random. 
This is consistent with the detailed models of the pulse profiles in several pulsars that show less than 10\degr\ displacement 
of the dipole \citep{1995MNRAS.277.1177L,1996ApJ...467..794K}. 
These small displacements will not affect our conclusions. 

\section{Conclusions}

\begin{enumerate}

\item We have collected pulse profiles of 124 X-ray pulsars and magnetars and 
classified them according to the number of pulses visible in one rotational period.
At energies above 10 keV, where the effects of photoelectric absorption and 
cyclotron line are minimal, 55\% of the pulsars have double-peaked profiles, while 
for all pulsars this fraction is 44\%. 

\item  We considered a simple   model with two point-like antipodal spots emitting radiation according to 
different types of pencil-beam patterns $\cos^{n}{\alpha}$ (where $n\ge 1$) and $\cos{\alpha}(1+h\cos{\alpha})$ (with $h\ge0$). 
The light curves produced by these cases are either single- or double-peaked depending on the model parameters. 

\item
We obtained some constraints on the radiation model parameters in the blackbody case, $n=1$ ($h=0$). 
The relative fraction of double-peaked profiles here depends on the neutron star compactness. 
The most probable values for the maximal gravitational light deflection angle is $\Deltag\sim 31\degr$ (for unconstrained magnetic field, $\theta_{\rm{m}}=\pi/2$ and with a full detectability of the secondary pulse). We also obtained a lower limit on the maximum 
 magnetic inclination to be $\thetamax\gtrsim 35\degr$, and for reasonable NS compactnesses we have $\thetamax\gtrsim 70\degr$.
When we include the effect of the detection threshold, the limit on the compactness is reduced to $\Deltag\lesssim 30\degr$ (at 95\% confidence) for $T=0.2$.  

\item
Any pencil-beam pattern (if not a blackbody) predicts a fixed  fraction of double-peaked profiles of 79\%, 
which is inconsistent with the data. 
Restricting the maximum magnetic inclination reduces the fraction of double-peaked profiles. 
Comparison to the data gives us the most probable value for the maximum magnetic inclination of $\thetamax= 40\degr\pm4\degr$. 
The neutron star compactness, however, cannot be constrained at all. 

\item A limited detection sensitivity to weak pulses also reduces the fraction of double-peaked profiles. 
We found that this fraction depends weakly on the neutron star compactness and is consistent with the data 
for a large range of pencil-beam patterns at $T\sim0.2$. 
In this case, we do not find good evidence that the magnetic inclination has a strict upper limit.

\item 
The overall conclusion is that contrary to the previous claims made by B03, 
the statistical method based on the classification of pulsar profiles by number of peaks
cannot constrain the compactness of the neutron star.
We also do not find univocal evidences in favor of the alignment of the magnetic dipole. 
It seems that the detailed analysis of the pulse profiles of individual pulsars and their evolution 
is the only way to obtain any useful constraints on the neutron star compactness and the magnetic field geometry.

\end{enumerate}

\begin{acknowledgements} 
This research was supported by Space institute, University of Oulu,
and the V\"ais\"ala foundation (AM).  JP acknowledges support from the
Academy of Finland grant 110792. We also acknowledge the support of
the International Space Science Institute (Bern, Switzerland), where
part of this investigation was carried out. We thank  Alexander Lutovinov 
and  the referee for the valuable comments. 
\end{acknowledgements}


\begin{thebibliography}{145}
\expandafter\ifx\csname natexlab\endcsname\relax\def\natexlab#1{#1}\fi

\bibitem[{{Angelini} {et~al.}(1998){Angelini}, {Church}, {Parmar},
  {Balucinska-Church}, \& {Mineo}}]{1998AAA...339L..41A}
{Angelini}, L., {Church}, M.~J., {Parmar}, A.~N., {Balucinska-Church}, M., \&
  {Mineo}, T. 1998, \aap, 339, L41

\bibitem[{{Aoki} {et~al.}(1992){Aoki}, {Dotani}, {Ebisawa}, {Itoh}, {Makino},
  {Nagase}, {Takeshima}, {Mihara}, \& {Kitamoto}}]{1992PASJ...44..641A}
{Aoki}, T., {Dotani}, T., {Ebisawa}, K., {et~al.} 1992, \pasj, 44, 641

\bibitem[{{Arons} \& {Lea}(1980)}]{1980ApJ...235.1016A}
{Arons}, J. \& {Lea}, S.~M. 1980, \apj, 235, 1016

\bibitem[{{Augello} {et~al.}(2003){Augello}, {Iaria}, {Robba}, {Di Salvo},
  {Burderi}, {Lavagetto}, \& {Stella}}]{2003ApJ...596L..63A}
{Augello}, G., {Iaria}, R., {Robba}, N.~R., {et~al.} 2003, \apjl, 596, L63

\bibitem[{{Bamba} {et~al.}(2001){Bamba}, {Yokogawa}, {Ueno}, {Koyama}, \&
  {Yamauchi}}]{2001PASJ...53.1179B}
{Bamba}, A., {Yokogawa}, J., {Ueno}, M., {Koyama}, K., \& {Yamauchi}, S. 2001,
  \pasj, 53, 1179

\bibitem[{{Basko} \& {Sunyaev}(1976)}]{1976MNRAS.175..395B}
{Basko}, M.~M. \& {Sunyaev}, R.~A. 1976, \mnras, 175, 395

\bibitem[{{Becker} \& {Wolff}(2005)}]{2005ApJ...621L..45B}
{Becker}, P.~A. \& {Wolff}, M.~T. 2005, \apjl, 621, L45

\bibitem[{{Beloborodov}(2002)}]{2002ApJ...566L..85B}
{Beloborodov}, A.~M. 2002, \apjl, 566, L85

\bibitem[{{Beloborodov} \& {Thompson}(2007)}]{2007ApJ...657..967B}
{Beloborodov}, A.~M. \& {Thompson}, C. 2007, \apj, 657, 967

\bibitem[{{Bildsten} {et~al.}(1997){Bildsten}, {Chakrabarty}, {Chiu}, {Finger},
  {Koh}, {Nelson}, {Prince}, {Rubin}, {Scott}, {Stollberg}, {Vaughan},
  {Wilson}, \& {Wilson}}]{1997ApJS..113..367B}
{Bildsten}, L., {Chakrabarty}, D., {Chiu}, J., {et~al.} 1997, \apjs, 113, 367

\bibitem[{{Bodaghee} {et~al.}(2006){Bodaghee}, {Walter}, {Zurita Heras},
  {Bird}, {Courvoisier}, {Malizia}, {Terrier}, \&
  {Ubertini}}]{2006AA...447.1027B}
{Bodaghee}, A., {Walter}, R., {Zurita Heras}, J.~A., {et~al.} 2006, \aap, 447,
  1027

\bibitem[{{Bulik} {et~al.}(2003){Bulik}, {Gondek-Rosi{\'n}ska}, {Santangelo},
  {Mihara}, {Finger}, \& {Cemeljic}}]{2003AAA...404.1023B}
{Bulik}, T., {Gondek-Rosi{\'n}ska}, D., {Santangelo}, A., {et~al.} 2003, \aap,
  404, 1023 (B03)

\bibitem[{{Bulik} {et~al.}(1995){Bulik}, {Riffert}, {Meszaros}, {Makishima},
  {Mihara}, \& {Thomas}}]{1995ApJ...444..405B}
{Bulik}, T., {Riffert}, H., {Meszaros}, P., {et~al.} 1995, \apj, 444, 405

\bibitem[{{Burderi} {et~al.}(1998){Burderi}, {di Salvo}, {Robba}, {del Sordo},
  {Santangelo}, \& {Segreto}}]{1998ApJ...498..831B}
{Burderi}, L., {di Salvo}, T., {Robba}, N.~R., {et~al.} 1998, \apj, 498, 831

\bibitem[{{Burderi} {et~al.}(2000){Burderi}, {Di Salvo}, {Robba}, {La Barbera},
  \& {Guainazzi}}]{2000ApJ...530..429B}
{Burderi}, L., {Di Salvo}, T., {Robba}, N.~R., {La Barbera}, A., \&
  {Guainazzi}, M. 2000, \apj, 530, 429

\bibitem[{{Burnard} {et~al.}(1991){Burnard}, {Arons}, \&
  {Klein}}]{1991ApJ...367..575B}
{Burnard}, D.~J., {Arons}, J., \& {Klein}, R.~I. 1991, \apj, 367, 575

\bibitem[{{Chakrabarty} {et~al.}(1995){Chakrabarty}, {Koh}, {Bildsten},
  {Prince}, {Finger}, {Wilson}, {Pendleton}, \& {Rubin}}]{1995ApJ...446..826C}
{Chakrabarty}, D., {Koh}, T., {Bildsten}, L., {et~al.} 1995, \apj, 446, 826

\bibitem[{{Chernyakova} {et~al.}(2005){Chernyakova}, {Lutovinov},
  {Rodr{\'{\i}}guez}, \& {Revnivtsev}}]{2005MNRAS.364..455C}
{Chernyakova}, M., {Lutovinov}, A., {Rodr{\'{\i}}guez}, J., \& {Revnivtsev}, M.
  2005, \mnras, 364, 455

\bibitem[{{Coe} {et~al.}(1994){Coe}, {Roche}, {Everall}, {Fishman}, {Hagedon},
  {Finger}, {Wilson}, {Buckley}, {Shrader}, {Fabregat}, {Polcaro},
  {Giovannelli}, \& {Villada}}]{1994AAA...289..784C}
{Coe}, M.~J., {Roche}, P., {Everall}, C., {et~al.} 1994, \aap, 289, 784

\bibitem[{{Corbet} {et~al.}(2001){Corbet}, {Marshall}, {Coe}, {Laycock}, \&
  {Handler}}]{2001ApJ...548L..41C}
{Corbet}, R.~H.~D., {Marshall}, F.~E., {Coe}, M.~J., {Laycock}, S., \&
  {Handler}, G. 2001, \apjl, 548, L41

\bibitem[{{Corbet} {et~al.}(1999){Corbet}, {Marshall}, {Peele}, \&
  {Takeshima}}]{1999ApJ...517..956C}
{Corbet}, R.~H.~D., {Marshall}, F.~E., {Peele}, A.~G., \& {Takeshima}, T. 1999,
  \apj, 517, 956

\bibitem[{{Corbet} \& {Peele}(1997)}]{1997ApJ...489L..83C}
{Corbet}, R.~H.~D. \& {Peele}, A.~G. 1997, \apjl, 489, L83

\bibitem[{{Cusumano} {et~al.}(1998){Cusumano}, {Israel}, {Mannucci}, {Masetti},
  {Mineo}, \& {Nicastro}}]{1998AAA...337..772C}
{Cusumano}, G., {Israel}, G.~L., {Mannucci}, F., {et~al.} 1998, \aap, 337, 772

\bibitem[{{Cusumano} {et~al.}(2000){Cusumano}, {Maccarone}, {Nicastro},
  {Sacco}, \& {Kaaret}}]{2000ApJ...528L..25C}
{Cusumano}, G., {Maccarone}, M.~C., {Nicastro}, L., {Sacco}, B., \& {Kaaret},
  P. 2000, \apjl, 528, L25

\bibitem[{{Davidson} \& {Ostriker}(1973)}]{1973ApJ...179..585D}
{Davidson}, K. \& {Ostriker}, J.~P. 1973, \apj, 179, 585

\bibitem[{{Edge} {et~al.}(2004){Edge}, {Coe}, {Galache}, {McBride}, {Corbet},
  {Markwardt}, \& {Laycock}}]{2004MNRAS.353.1286E}
{Edge}, W.~R.~T., {Coe}, M.~J., {Galache}, J.~L., {et~al.} 2004, \mnras, 353,
  1286

\bibitem[{{Finger} {et~al.}(1999){Finger}, {Bildsten}, {Chakrabarty}, {Prince},
  {Scott}, {Wilson}, {Wilson}, \& {Zhang}}]{1999ApJ...517..449F}
{Finger}, M.~H., {Bildsten}, L., {Chakrabarty}, D., {et~al.} 1999, \apj, 517,
  449

\bibitem[{{Finger} {et~al.}(2001){Finger}, {Macomb}, {Lamb}, {Prince}, {Coe},
  \& {Haigh}}]{2001ApJ...560..378F}
{Finger}, M.~H., {Macomb}, D.~J., {Lamb}, R.~C., {et~al.} 2001, \apj, 560, 378

\bibitem[{{Galloway} {et~al.}(2005){Galloway}, {Wang}, \&
  {Morgan}}]{2005ApJ...635.1217G}
{Galloway}, D.~K., {Wang}, Z., \& {Morgan}, E.~H. 2005, \apj, 635, 1217

\bibitem[{{Giacconi} {et~al.}(1971){Giacconi}, {Gursky}, {Kellogg}, {Schreier},
  \& {Tananbaum}}]{1971ApJ...167L..67G}
{Giacconi}, R., {Gursky}, H., {Kellogg}, E., {Schreier}, E., \& {Tananbaum}, H.
  1971, \apjl, 167, L67

\bibitem[{{G{\"o}{\u g}{\"u}{\c s}} {et~al.}(2002){G{\"o}{\u g}{\"u}{\c s}},
  {Kouveliotou}, {Woods}, {Finger}, \& {van der Klis}}]{2002ApJ...577..929G}
{G{\"o}{\u g}{\"u}{\c s}}, E., {Kouveliotou}, C., {Woods}, P.~M., {Finger},
  M.~H., \& {van der Klis}, M. 2002, \apj, 577, 929

\bibitem[{{Haberl} {et~al.}(2003){Haberl}, {Dennerl}, \&
  {Pietsch}}]{2003AA...406..471H}
{Haberl}, F., {Dennerl}, K., \& {Pietsch}, W. 2003, \aap, 406, 471

\bibitem[{{Haberl} {et~al.}(1997){Haberl}, {Dennerl}, {Pietsch}, \&
  {Reinsch}}]{1997AAA...318..490H}
{Haberl}, F., {Dennerl}, K., {Pietsch}, W., \& {Reinsch}, K. 1997, \aap, 318,
  490

\bibitem[{{Haberl} \& {Pietsch}(2005)}]{2005AA...438..211H}
{Haberl}, F. \& {Pietsch}, W. 2005, \aap, 438, 211

\bibitem[{{Haberl} \& {Pietsch}(2008)}]{2008AA...484..451H}
{Haberl}, F. \& {Pietsch}, W. 2008, \aap, 484, 451

\bibitem[{{Haberl} {et~al.}(2004{\natexlab{a}}){Haberl}, {Pietsch}, {Schartel},
  {Rodriguez}, \& {Corbet}}]{2004AA...420L..19H}
{Haberl}, F., {Pietsch}, W., {Schartel}, N., {Rodriguez}, P., \& {Corbet},
  R.~H.~D. 2004{\natexlab{a}}, \aap, 420, L19

\bibitem[{{Haberl} {et~al.}(2004{\natexlab{b}}){Haberl}, {Zavlin},
  {Tr{\"u}mper}, \& {Burwitz}}]{2004AAA...419.1077H}
{Haberl}, F., {Zavlin}, V.~E., {Tr{\"u}mper}, J., \& {Burwitz}, V.
  2004{\natexlab{b}}, \aap, 419, 1077

\bibitem[{{Haensel} {et~al.}(2007){Haensel}, {Potekhin}, \& {Yakovlev}}]{HPY07}
{Haensel}, P., {Potekhin}, A.~Y., \& {Yakovlev}, D.~G. 2007, Astrophysics and
  Space Science Library, Vol. 326, {Neutron Stars 1: Equation of State and
  Structure} (New York: Springer)

\bibitem[{{Hall} {et~al.}(2000){Hall}, {Finley}, {Corbet}, \&
  {Thomas}}]{2000ApJ...536..450H}
{Hall}, T.~A., {Finley}, J.~P., {Corbet}, R.~H.~D., \& {Thomas}, R.~C. 2000,
  \apj, 536, 450

\bibitem[{{Halpern} \& {Gotthelf}(2005)}]{2005ApJ...618..874H}
{Halpern}, J.~P. \& {Gotthelf}, E.~V. 2005, \apj, 618, 874

\bibitem[{{Halpern} \& {Gotthelf}(2007)}]{2007ApJ...669..579H}
{Halpern}, J.~P. \& {Gotthelf}, E.~V. 2007, \apj, 669, 579

\bibitem[{{Hulleman} {et~al.}(1998){Hulleman}, {in 't Zand}, \&
  {Heise}}]{1998AAA...337L..25H}
{Hulleman}, F., {in 't Zand}, J.~J.~M., \& {Heise}, J. 1998, \aap, 337, L25

\bibitem[{{Ibragimov} \& {Poutanen}(2009)}]{IP09}
{Ibragimov}, A. \& {Poutanen}, J. 2009, \mnras, 400, 492

\bibitem[{{Imanishi} {et~al.}(1999){Imanishi}, {Yokogawa}, {Tsujimoto}, \&
  {Koyama}}]{1999PASJ...51L..15I}
{Imanishi}, K., {Yokogawa}, J., {Tsujimoto}, M., \& {Koyama}, K. 1999, \pasj,
  51, L15

\bibitem[{{in 't Zand} {et~al.}(1998){in 't Zand}, {Baykal}, \&
  {Strohmayer}}]{1998ApJ...496..386I}
{in 't Zand}, J.~J.~M., {Baykal}, A., \& {Strohmayer}, T.~E. 1998, \apj, 496,
  386

\bibitem[{{in't Zand} \& {Heise}(2004)}]{2004ATel..362....1I}
{in't Zand}, J. \& {Heise}, J. 2004, The Astronomer's Telegram, 362, 1

\bibitem[{{in't Zand} {et~al.}(2001{\natexlab{a}}){in't Zand}, {Corbet}, \&
  {Marshall}}]{2001ApJ...553L.165I}
{in't Zand}, J.~J.~M., {Corbet}, R.~H.~D., \& {Marshall}, F.~E.
  2001{\natexlab{a}}, \apjl, 553, L165

\bibitem[{{in't Zand} {et~al.}(2001{\natexlab{b}}){in't Zand}, {Swank},
  {Corbet}, \& {Markwardt}}]{2001AA...380L..26I}
{in't Zand}, J.~J.~M., {Swank}, J., {Corbet}, R.~H.~D., \& {Markwardt}, C.~B.
  2001{\natexlab{b}}, \aap, 380, L26

\bibitem[{{Israel} {et~al.}(2000){Israel}, {Campana}, {Covino}, {Dal Fiume},
  {Gaetz}, {Mereghetti}, {Oosterbroek}, {Orlandini}, {Parmar}, {Ricci}, \&
  {Stella}}]{2000ApJ...531L.131I}
{Israel}, G.~L., {Campana}, S., {Covino}, S., {et~al.} 2000, \apjl, 531, L131

\bibitem[{{Israel} {et~al.}(2007){Israel}, {Campana}, {Dall'Osso}, {Muno},
  {Cummings}, {Perna}, \& {Stella}}]{2007ApJ...664..448I}
{Israel}, G.~L., {Campana}, S., {Dall'Osso}, S., {et~al.} 2007, \apj, 664, 448

\bibitem[{{Israel} {et~al.}(1997{\natexlab{a}}){Israel}, {Stella}, {Angelini},
  {White}, {Giommi}, \& {Covino}}]{1997ApJ...484L.141I}
{Israel}, G.~L., {Stella}, L., {Angelini}, L., {et~al.} 1997{\natexlab{a}},
  \apjl, 484, L141

\bibitem[{{Israel} {et~al.}(1997{\natexlab{b}}){Israel}, {Stella}, {Angelini},
  {White}, {Kallman}, {Giommi}, \& {Treves}}]{1997ApJ...474L..53I}
{Israel}, G.~L., {Stella}, L., {Angelini}, L., {et~al.} 1997{\natexlab{b}},
  \apjl, 474, L53

\bibitem[{{Iwasawa} {et~al.}(1992){Iwasawa}, {Koyama}, \&
  {Halpern}}]{1992PASJ...44....9I}
{Iwasawa}, K., {Koyama}, K., \& {Halpern}, J.~P. 1992, \pasj, 44, 9

\bibitem[{{Karasev} {et~al.}(2008){Karasev}, {Tsygankov}, \&
  {Lutovinov}}]{2008MNRAS.386L..10K}
{Karasev}, D.~I., {Tsygankov}, S.~S., \& {Lutovinov}, A.~A. 2008, \mnras, 386,
  L10

\bibitem[{{Kelley} {et~al.}(1983){Kelley}, {Rappaport}, \&
  {Ayasli}}]{1983ApJ...274..765K}
{Kelley}, R.~L., {Rappaport}, S., \& {Ayasli}, S. 1983, \apj, 274, 765

\bibitem[{{Kinugasa} {et~al.}(1998){Kinugasa}, {Torii}, {Hashimoto}, {Tsunemi},
  {Hayashida}, {Kitamoto}, {Kamata}, {Dotani}, {Nagase}, {Sugizaki}, {Ueda},
  {Kawai}, {Makishima}, \& {Yamauchi}}]{1998ApJ...495..435K}
{Kinugasa}, K., {Torii}, K., {Hashimoto}, Y., {et~al.} 1998, \apj, 495, 435

\bibitem[{{Kirk} {et~al.}(1986){Kirk}, {Nagel}, \&
  {Storey}}]{1986A&A...169..259K}
{Kirk}, J.~G., {Nagel}, W., \& {Storey}, M.~C. 1986, \aap, 169, 259

\bibitem[{{Koh} {et~al.}(1997){Koh}, {Bildsten}, {Chakrabarty}, {Nelson},
  {Prince}, {Vaughan}, {Finger}, {Wilson}, \& {Rubin}}]{1997ApJ...479..933K}
{Koh}, D.~T., {Bildsten}, L., {Chakrabarty}, D., {et~al.} 1997, \apj, 479, 933

\bibitem[{{Kohno} {et~al.}(2000){Kohno}, {Yokogawa}, \&
  {Koyama}}]{2000PASJ...52..299K}
{Kohno}, M., {Yokogawa}, J., \& {Koyama}, K. 2000, \pasj, 52, 299

\bibitem[{{Koyama} {et~al.}(1990){Koyama}, {Kawada}, {Takeuchi}, {Tawara},
  {Ushimaru}, {Dotani}, \& {Takizawa}}]{1990ApJ...356L..47K}
{Koyama}, K., {Kawada}, M., {Takeuchi}, Y., {et~al.} 1990, \apjl, 356, L47

\bibitem[{{Koyama} {et~al.}(1991{\natexlab{a}}){Koyama}, {Kawada}, {Tawara},
  {Kimura}, {Kitamoto}, {Miyamoto}, {Tsunemi}, {Ebisawa}, \&
  {Nagase}}]{1991ApJ...366L..19K}
{Koyama}, K., {Kawada}, M., {Tawara}, Y., {et~al.} 1991{\natexlab{a}}, \apjl,
  366, L19

\bibitem[{{Koyama} {et~al.}(1991{\natexlab{b}}){Koyama}, {Kunieda}, {Takeuchi},
  \& {Tawara}}]{1991ApJ...370L..77K}
{Koyama}, K., {Kunieda}, H., {Takeuchi}, Y., \& {Tawara}, Y.
  1991{\natexlab{b}}, \apjl, 370, L77

\bibitem[{{Kraus}(2001)}]{2001ApJ...563..289K}
{Kraus}, U. 2001, \apj, 563, 289

\bibitem[{{Kraus} {et~al.}(1996){Kraus}, {Blum}, {Schulte}, {Ruder}, \&
  {Meszaros}}]{1996ApJ...467..794K}
{Kraus}, U., {Blum}, S., {Schulte}, J., {Ruder}, H., \& {Meszaros}, P. 1996,
  \apj, 467, 794

\bibitem[{{Kraus} {et~al.}(2003){Kraus}, {Zahn}, {Weth}, \&
  {Ruder}}]{2003ApJ...590..424K}
{Kraus}, U., {Zahn}, C., {Weth}, C., \& {Ruder}, H. 2003, \apj, 590, 424

\bibitem[{{Kuiper} {et~al.}(2006){Kuiper}, {Hermsen}, {den Hartog}, \&
  {Collmar}}]{2006ApJ...645..556K}
{Kuiper}, L., {Hermsen}, W., {den Hartog}, P.~R., \& {Collmar}, W. 2006, \apj,
  645, 556

\bibitem[{{Kuiper} {et~al.}(2004){Kuiper}, {Hermsen}, \&
  {Mendez}}]{2004ApJ...613.1173K}
{Kuiper}, L., {Hermsen}, W., \& {Mendez}, M. 2004, \apj, 613, 1173

\bibitem[{{Lamb} {et~al.}(2002){Lamb}, {Macomb}, {Prince}, \&
  {Majid}}]{2002ApJ...567L.129L}
{Lamb}, R.~C., {Macomb}, D.~J., {Prince}, T.~A., \& {Majid}, W.~A. 2002, \apjl,
  567, L129

\bibitem[{{Laycock} {et~al.}(2003){Laycock}, {Corbet}, {Coe}, {Marshall},
  {Markwardt}, \& {Edge}}]{2003MNRAS.339..435L}
{Laycock}, S., {Corbet}, R.~H.~D., {Coe}, M.~J., {et~al.} 2003, \mnras, 339,
  435

\bibitem[{{Laycock} {et~al.}(2005){Laycock}, {Corbet}, {Coe}, {Marshall},
  {Markwardt}, \& {Lochner}}]{2005ApJS..161...96L}
{Laycock}, S., {Corbet}, R.~H.~D., {Coe}, M.~J., {et~al.} 2005, \apjs, 161, 96

\bibitem[{{Laycock} {et~al.}(2002){Laycock}, {Corbet}, {Perrodin}, {Coe},
  {Marshall}, \& {Markwardt}}]{2002AA...385..464L}
{Laycock}, S., {Corbet}, R.~H.~D., {Perrodin}, D., {et~al.} 2002, \aap, 385,
  464

\bibitem[{{Leahy}(1990)}]{1990MNRAS.242..188L}
{Leahy}, D.~A. 1990, \mnras, 242, 188

\bibitem[{{Leahy}(2003)}]{2003ApJ...596.1131L}
{Leahy}, D.~A. 2003, \apj, 596, 1131

\bibitem[{{Leahy} \& {Li}(1995)}]{1995MNRAS.277.1177L}
{Leahy}, D.~A. \& {Li}, L. 1995, \mnras, 277, 1177

\bibitem[{{Levine} {et~al.}(1993){Levine}, {Rappaport}, {Deeter}, {Boynton}, \&
  {Nagase}}]{1993ApJ...410..328L}
{Levine}, A., {Rappaport}, S., {Deeter}, J.~E., {Boynton}, P.~E., \& {Nagase},
  F. 1993, \apj, 410, 328

\bibitem[{{Levine} {et~al.}(2004){Levine}, {Rappaport}, {Remillard}, \&
  {Savcheva}}]{2004ApJ...617.1284L}
{Levine}, A.~M., {Rappaport}, S., {Remillard}, R., \& {Savcheva}, A. 2004,
  \apj, 617, 1284

\bibitem[{{Lutovinov} {et~al.}(2005{\natexlab{a}}){Lutovinov}, {Revnivtsev},
  {Gilfanov}, {Shtykovskiy}, {Molkov}, \& {Sunyaev}}]{2005AA...444..821L}
{Lutovinov}, A., {Revnivtsev}, M., {Gilfanov}, M., {et~al.} 2005{\natexlab{a}},
  \aap, 444, 821

\bibitem[{{Lutovinov} {et~al.}(2005{\natexlab{b}}){Lutovinov}, {Rodriguez},
  {Revnivtsev}, \& {Shtykovskiy}}]{2005AA...433L..41L}
{Lutovinov}, A., {Rodriguez}, J., {Revnivtsev}, M., \& {Shtykovskiy}, P.
  2005{\natexlab{b}}, \aap, 433, L41

\bibitem[{{Macomb} {et~al.}(1999){Macomb}, {Finger}, {Harmon}, {Lamb}, \&
  {Prince}}]{1999ApJ...518L..99M}
{Macomb}, D.~J., {Finger}, M.~H., {Harmon}, B.~A., {Lamb}, R.~C., \& {Prince},
  T.~A. 1999, \apjl, 518, L99

\bibitem[{{Macomb} {et~al.}(2003){Macomb}, {Fox}, {Lamb}, \&
  {Prince}}]{2003ApJ...584L..79M}
{Macomb}, D.~J., {Fox}, D.~W., {Lamb}, R.~C., \& {Prince}, T.~A. 2003, \apjl,
  584, L79

\bibitem[{{Majid} {et~al.}(2004){Majid}, {Lamb}, \&
  {Macomb}}]{2004ApJ...609..133M}
{Majid}, W.~A., {Lamb}, R.~C., \& {Macomb}, D.~J. 2004, \apj, 609, 133

\bibitem[{{McBride} {et~al.}(2006){McBride}, {Wilms}, {Coe}, {Kreykenbohm},
  {Rothschild}, {Coburn}, {Galache}, {Kretschmar}, {Edge}, \&
  {Staubert}}]{2006AA...451..267M}
{McBride}, V.~A., {Wilms}, J., {Coe}, M.~J., {et~al.} 2006, \aap, 451, 267

\bibitem[{{McClintock} {et~al.}(1977){McClintock}, {Nugent}, {Li}, \&
  {Rappaport}}]{1977ApJ...216L..15M}
{McClintock}, J.~E., {Nugent}, J.~J., {Li}, F.~K., \& {Rappaport}, S.~A. 1977,
  \apjl, 216, L15

\bibitem[{{Mereghetti}(2008)}]{2008A&ARv..15..225M}
{Mereghetti}, S. 2008, \aapr, 15, 225

\bibitem[{{Mereghetti} {et~al.}(2002){Mereghetti}, {Chiarlone}, {Israel}, \&
  {Stella}}]{2002nsps.conf...29M}
{Mereghetti}, S., {Chiarlone}, L., {Israel}, G.~L., \& {Stella}, L. 2002, in
  Neutron Stars, Pulsars, and Supernova Remnants, ed. {W.~Becker, H.~Lesch, \&
  J.~Tr{\"u}mper}, 29

\bibitem[{{Mereghetti} {et~al.}(2005){Mereghetti}, {G{\"o}tz}, {Mirabel}, \&
  {Hurley}}]{2005A&A...433L...9M}
{Mereghetti}, S., {G{\"o}tz}, D., {Mirabel}, I.~F., \& {Hurley}, K. 2005, \aap,
  433, L9

\bibitem[{{Mereghetti} {et~al.}(2000){Mereghetti}, {Tiengo}, {Israel}, \&
  {Stella}}]{2000AAA...354..567M}
{Mereghetti}, S., {Tiengo}, A., {Israel}, G.~L., \& {Stella}, L. 2000, \aap,
  354, 567

\bibitem[{{Meszaros} \& {Nagel}(1985{\natexlab{a}})}]{1985ApJ...298..147M}
{Meszaros}, P. \& {Nagel}, W. 1985{\natexlab{a}}, \apj, 298, 147

\bibitem[{{Meszaros} \& {Nagel}(1985{\natexlab{b}})}]{1985ApJ...299..138M}
{Meszaros}, P. \& {Nagel}, W. 1985{\natexlab{b}}, \apj, 299, 138

\bibitem[{{Mihara}(1995)}]{1995PhDT.......215M}
{Mihara}, T. 1995, PhD thesis, Dept.~of Physics, Univ.~of Tokyo

\bibitem[{{Molkov} {et~al.}(2005){Molkov}, {Hurley}, {Sunyaev}, {Shtykovsky},
  {Revnivtsev}, \& {Kouveliotou}}]{2005A&A...433L..13M}
{Molkov}, S., {Hurley}, K., {Sunyaev}, R., {et~al.} 2005, \aap, 433, L13

\bibitem[{{Morii} {et~al.}(2003){Morii}, {Sato}, {Kataoka}, \&
  {Kawai}}]{2003PASJ...55L..45M}
{Morii}, M., {Sato}, R., {Kataoka}, J., \& {Kawai}, N. 2003, \pasj, 55, L45

\bibitem[{{Morris} {et~al.}(2009){Morris}, {Smith}, {Markwardt}, {Mushotzky},
  {Tueller}, {Kallman}, \& {Dhuga}}]{2009ApJ...699..892M}
{Morris}, D.~C., {Smith}, R.~K., {Markwardt}, C.~B., {et~al.} 2009, \apj, 699,
  892

\bibitem[{{Nagase}(1989)}]{1989PASJ...41....1N}
{Nagase}, F. 1989, \pasj, 41, 1

\bibitem[{{Nagel}(1981{\natexlab{a}})}]{1981ApJ...251..288N}
{Nagel}, W. 1981{\natexlab{a}}, \apj, 251, 288

\bibitem[{{Nagel}(1981{\natexlab{b}})}]{1981ApJ...251..278N}
{Nagel}, W. 1981{\natexlab{b}}, \apj, 251, 278

\bibitem[{{Negueruela} {et~al.}(2000){Negueruela}, {Reig}, \&
  {Clark}}]{2000A&A...354L..29N}
{Negueruela}, I., {Reig}, P., \& {Clark}, J.~S. 2000, \aap, 354, L29

\bibitem[{{Oosterbroek} {et~al.}(1999){Oosterbroek}, {Orlandini}, {Parmar},
  {Angelini}, {Israel}, {Dal Fiume}, {Mereghetti}, {Santangelo}, \&
  {Cusumano}}]{1999AAA...351L..33O}
{Oosterbroek}, T., {Orlandini}, M., {Parmar}, A.~N., {et~al.} 1999, \aap, 351,
  L33

\bibitem[{{Paul} \& {Rao}(1998)}]{1998A&A...337..815P}
{Paul}, B. \& {Rao}, A.~R. 1998, \aap, 337, 815

\bibitem[{{Pechenick} {et~al.}(1983){Pechenick}, {Ftaclas}, \&
  {Cohen}}]{1983ApJ...274..846P}
{Pechenick}, K.~R., {Ftaclas}, C., \& {Cohen}, J.~M. 1983, \apj, 274, 846

\bibitem[{{Poutanen}(2008)}]{P08AIP}
{Poutanen}, J. 2008, in A Decade of Accreting Millisecond X-ray Pulsars, 
ed. {R.~Wijnands,  D.~Altamirano, P.~Soleri, N.~Degenaar, N.~Rea, P.~Casella, A.~Patruno, \&
  M.~Linares} (New York: AIP), AIP Conf. Ser. 1068,  77

\bibitem[{{Poutanen} \& {Beloborodov}(2006)}]{PB06}
{Poutanen}, J. \& {Beloborodov}, A.~M. 2006, \mnras, 373, 836

\bibitem[{{Poutanen} \& {Gierli{\'n}ski}(2003)}]{PG03}
{Poutanen}, J. \& {Gierli{\'n}ski}, M. 2003, \mnras, 343, 1301

\bibitem[{{Reig} {et~al.}(2008){Reig}, {Belloni}, {Israel}, {Campana},
  {Gehrels}, \& {Homan}}]{2008AA...485..797R}
{Reig}, P., {Belloni}, T., {Israel}, G.~L., {et~al.} 2008, \aap, 485, 797

\bibitem[{{Reig} \& {Roche}(1999{\natexlab{a}})}]{1999MNRAS.306..100R}
{Reig}, P. \& {Roche}, P. 1999{\natexlab{a}}, \mnras, 306, 100

\bibitem[{{Reig} \& {Roche}(1999{\natexlab{b}})}]{1999MNRAS.306...95R}
{Reig}, P. \& {Roche}, P. 1999{\natexlab{b}}, \mnras, 306, 95

\bibitem[{{Robba} \& {Warwick}(1989)}]{1989ApJ...346..469R}
{Robba}, N.~R. \& {Warwick}, R.~S. 1989, \apj, 346, 469

\bibitem[{{Sakano} {et~al.}(2000){Sakano}, {Torii}, {Koyama}, {Maeda}, \&
  {Yamauchi}}]{2000PASJ...52.1141S}
{Sakano}, M., {Torii}, K., {Koyama}, K., {Maeda}, Y., \& {Yamauchi}, S. 2000,
  \pasj, 52, 1141

\bibitem[{{Santangelo} {et~al.}(1998){Santangelo}, {Cusumano}, {dal Fiume},
  {Israel}, {Stella}, {Orlandini}, \& {Parmar}}]{1998AAA...338L..59S}
{Santangelo}, A., {Cusumano}, G., {dal Fiume}, D., {et~al.} 1998, \aap, 338,
  L59

\bibitem[{{Santangelo} {et~al.}(1999){Santangelo}, {Segreto}, {Giarrusso}, {dal
  Fiume}, {Orlandini}, {Parmar}, {Oosterbroek}, {Bulik}, {Mihara}, {Campana},
  {Israel}, \& {Stella}}]{1999ApJ...523L..85S}
{Santangelo}, A., {Segreto}, A., {Giarrusso}, S., {et~al.} 1999, \apjl, 523,
  L85

\bibitem[{{Sasaki} {et~al.}(2003){Sasaki}, {Pietsch}, \&
  {Haberl}}]{2003AA...403..901S}
{Sasaki}, M., {Pietsch}, W., \& {Haberl}, F. 2003, \aap, 403, 901

\bibitem[{{Schmidtke} {et~al.}(1995){Schmidtke}, {Cowley}, {McGrath}, \&
  {Anderson}}]{1995PASP..107..450S}
{Schmidtke}, P.~C., {Cowley}, A.~P., {McGrath}, T.~K., \& {Anderson}, A.~L.
  1995, \pasp, 107, 450

\bibitem[{{Scott} {et~al.}(1997){Scott}, {Finger}, {Wilson}, {Koh}, {Prince},
  {Vaughan}, \& {Chakrabarty}}]{1997ApJ...488..831S}
{Scott}, D.~M., {Finger}, M.~H., {Wilson}, R.~B., {et~al.} 1997, \apj, 488, 831

\bibitem[{{Seward} {et~al.}(1986){Seward}, {Charles}, \&
  {Smale}}]{1986ApJ...305..814S}
{Seward}, F.~D., {Charles}, P.~A., \& {Smale}, A.~P. 1986, \apj, 305, 814

\bibitem[{{Sguera} {et~al.}(2007){Sguera}, {Hill}, {Bird}, {Dean}, {Bazzano},
  {Ubertini}, {Masetti}, {Landi}, {Malizia}, {Clark}, \&
  {Molina}}]{2007AA...467..249S}
{Sguera}, V., {Hill}, A.~B., {Bird}, A.~J., {et~al.} 2007, \aap, 467, 249

\bibitem[{{Sidoli} {et~al.}(2007){Sidoli}, {Romano}, {Mereghetti}, {Paizis},
  {Vercellone}, {Mangano}, \& {G{\"o}tz}}]{2007AA...476.1307S}
{Sidoli}, L., {Romano}, P., {Mereghetti}, S., {et~al.} 2007, \aap, 476, 1307

\bibitem[{{Skinner} {et~al.}(1982){Skinner}, {Bedford}, {Elsner}, {Leahy},
  {Weisskopf}, \& {Grindlay}}]{1982Natur.297..568S}
{Skinner}, G.~K., {Bedford}, D.~K., {Elsner}, R.~F., {et~al.} 1982, \nat, 297,
  568

\bibitem[{{Stollberg} {et~al.}(1999){Stollberg}, {Finger}, {Wilson}, {Scott},
  {Crary}, \& {Paciesas}}]{1999ApJ...512..313S}
{Stollberg}, M.~T., {Finger}, M.~H., {Wilson}, R.~B., {et~al.} 1999, \apj, 512,
  313

\bibitem[{{Suleimanov} \& {Poutanen}(2006)}]{2006MNRAS.369.2036S}
{Suleimanov}, V. \& {Poutanen}, J. 2006, \mnras, 369, 2036

\bibitem[{{Tawara} {et~al.}(1989){Tawara}, {Yamauchi}, {Awaki}, {Kii},
  {Koyama}, \& {Nagase}}]{1989PASJ...41..473T}
{Tawara}, Y., {Yamauchi}, S., {Awaki}, H., {et~al.} 1989, \pasj, 41, 473

\bibitem[{{Thompson} \& {Beloborodov}(2005)}]{2005ApJ...634..565T}
{Thompson}, C. \& {Beloborodov}, A.~M. 2005, \apj, 634, 565

\bibitem[{{Thorsett} \& {Chakrabarty}(1999)}]{1999ApJ...512..288T}
{Thorsett}, S.~E. \& {Chakrabarty}, D. 1999, \apj, 512, 288

\bibitem[{{Torii} {et~al.}(1998{\natexlab{a}}){Torii}, {Kinugasa}, {Katayama},
  {Kohmura}, {Tsunemi}, {Sakano}, {Nishiuchi}, {Koyama}, \&
  {Yamauchi}}]{1998ApJ...508..854T}
{Torii}, K., {Kinugasa}, K., {Katayama}, K., {et~al.} 1998{\natexlab{a}}, \apj,
  508, 854

\bibitem[{{Torii} {et~al.}(1998{\natexlab{b}}){Torii}, {Kinugasa}, {Katayama},
  {Tsunemi}, \& {Yamauchi}}]{1998ApJ...503..843T}
{Torii}, K., {Kinugasa}, K., {Katayama}, K., {Tsunemi}, H., \& {Yamauchi}, S.
  1998{\natexlab{b}}, \apj, 503, 843

\bibitem[{{Torii} {et~al.}(1999){Torii}, {Sugizaki}, {Kohmura}, {Endo}, \&
  {Nagase}}]{1999ApJ...523L..65T}
{Torii}, K., {Sugizaki}, M., {Kohmura}, T., {Endo}, T., \& {Nagase}, F. 1999,
  \apjl, 523, L65

\bibitem[{{Trudolyubov} {et~al.}(2005){Trudolyubov}, {Kotov}, {Priedhorsky},
  {Cordova}, \& {Mason}}]{2005ApJ...634..314T}
{Trudolyubov}, S., {Kotov}, O., {Priedhorsky}, W., {Cordova}, F., \& {Mason},
  K. 2005, \apj, 634, 314

\bibitem[{{Trudolyubov}(2008)}]{2008MNRAS.387L..36T}
{Trudolyubov}, S.~P. 2008, \mnras, 387, L36

\bibitem[{{Trudolyubov} {et~al.}(2007){Trudolyubov}, {Priedhorsky}, \&
  {C{\'o}rdova}}]{2007ApJ...663..487T}
{Trudolyubov}, S.~P., {Priedhorsky}, W.~C., \& {C{\'o}rdova}, F.~A. 2007, \apj,
  663, 487

\bibitem[{{Tsujimoto} {et~al.}(1999){Tsujimoto}, {Imanishi}, {Yokogawa}, \&
  {Koyama}}]{1999PASJ...51L..21T}
{Tsujimoto}, M., {Imanishi}, K., {Yokogawa}, J., \& {Koyama}, K. 1999, \pasj,
  51, L21

\bibitem[{{Tsygankov} \& {Lutovinov}(2005)}]{2005AstL...31...88T}
{Tsygankov}, S.~S. \& {Lutovinov}, A.~A. 2005, Astr. Lett., 31, 88

\bibitem[{{Tsygankov} {et~al.}(2006){Tsygankov}, {Lutovinov}, {Churazov}, \&
  {Sunyaev}}]{2006MNRAS.371...19T}
{Tsygankov}, S.~S., {Lutovinov}, A.~A., {Churazov}, E.~M., \& {Sunyaev}, R.~A.
  2006, \mnras, 371, 19

\bibitem[{{Ueno} {et~al.}(2000){Ueno}, {Yokogawa}, {Imanishi}, \&
  {Koyama}}]{2000PASJ...52L..63U}
{Ueno}, M., {Yokogawa}, J., {Imanishi}, K., \& {Koyama}, K. 2000, \pasj, 52,
  L63

\bibitem[{{Viironen} \& {Poutanen}(2004)}]{VP04}
{Viironen}, K. \& {Poutanen}, J. 2004, \aap, 426, 985

\bibitem[{{Wang} \& {Welter}(1981)}]{1981AAA...102...97W}
{Wang}, Y.-M. \& {Welter}, G.~L. 1981, \aap, 102, 97

\bibitem[{{Wilson} {et~al.}(2003){Wilson}, {Finger}, {Coe}, \&
  {Negueruela}}]{2003ApJ...584..996W}
{Wilson}, C.~A., {Finger}, M.~H., {Coe}, M.~J., \& {Negueruela}, I. 2003, \apj,
  584, 996

\bibitem[{{Wilson} {et~al.}(2002){Wilson}, {Finger}, {G{\"o}{\u g}{\"u}{\c s}},
  {Woods}, \& {Kouveliotou}}]{2002ApJ...565.1150W}
{Wilson}, C.~A., {Finger}, M.~H., {G{\"o}{\u g}{\"u}{\c s}}, E., {Woods},
  P.~M., \& {Kouveliotou}, C. 2002, \apj, 565, 1150

\bibitem[{{Woo} {et~al.}(1996){Woo}, {Clark}, {Levine}, {Corbet}, \&
  {Nagase}}]{1996ApJ...467..811W}
{Woo}, J.~W., {Clark}, G.~W., {Levine}, A.~M., {Corbet}, R.~H.~D., \& {Nagase},
  F. 1996, \apj, 467, 811

\bibitem[{{Yahel}(1980)}]{1980A&A....90...26Y}
{Yahel}, R.~Z. 1980, \aap, 90, 26

\bibitem[{{Yokogawa} {et~al.}(1999){Yokogawa}, {Imanishi}, {Tsujimoto},
  {Kohno}, \& {Koyama}}]{1999PASJ...51..547Y}
{Yokogawa}, J., {Imanishi}, K., {Tsujimoto}, M., {Kohno}, M., \& {Koyama}, K.
  1999, \pasj, 51, 547

\bibitem[{{Yokogawa} {et~al.}(2000{\natexlab{a}}){Yokogawa}, {Imanishi},
  {Tsujimoto}, {Nishiuchi}, {Koyama}, {Nagase}, \&
  {Corbet}}]{2000ApJS..128..491Y}
{Yokogawa}, J., {Imanishi}, K., {Tsujimoto}, M., {et~al.} 2000{\natexlab{a}},
  \apjs, 128, 491

\bibitem[{{Yokogawa} {et~al.}(2000{\natexlab{b}}){Yokogawa}, {Imanishi},
  {Ueno}, \& {Koyama}}]{2000PASJ...52L..73Y}
{Yokogawa}, J., {Imanishi}, K., {Ueno}, M., \& {Koyama}, K. 2000{\natexlab{b}},
  \pasj, 52, L73

\bibitem[{{Yokogawa} {et~al.}(2000{\natexlab{c}}){Yokogawa}, {Paul}, {Ozaki},
  {Nagase}, {Chakrabarty}, \& {Takeshima}}]{2000ApJ...539..191Y}
{Yokogawa}, J., {Paul}, B., {Ozaki}, M., {et~al.} 2000{\natexlab{c}}, \apj,
  539, 191

\bibitem[{{Yokogawa} {et~al.}(2000{\natexlab{d}}){Yokogawa}, {Torii},
  {Imanishi}, \& {Koyama}}]{2000PASJ...52L..37Y}
{Yokogawa}, J., {Torii}, K., {Imanishi}, K., \& {Koyama}, K.
  2000{\natexlab{d}}, \pasj, 52, L37

\bibitem[{{Yokogawa} {et~al.}(2000{\natexlab{e}}){Yokogawa}, {Torii},
  {Kohmura}, {Imanishi}, \& {Koyama}}]{2000PASJ...52L..53Y}
{Yokogawa}, J., {Torii}, K., {Kohmura}, T., {Imanishi}, K., \& {Koyama}, K.
  2000{\natexlab{e}}, \pasj, 52, L53

\bibitem[{{Yokogawa} {et~al.}(2001){Yokogawa}, {Torii}, {Kohmura}, \&
  {Koyama}}]{2001PASJ...53L...9Y}
{Yokogawa}, J., {Torii}, K., {Kohmura}, T., \& {Koyama}, K. 2001, \pasj, 53, L9

\end{thebibliography}

\end{document}